\documentclass[conf]{new-aiaa}
\usepackage[utf8]{inputenc}

\usepackage{graphicx}
\usepackage{amsmath}
\usepackage{amssymb}
\usepackage[version=4]{mhchem}
\usepackage{siunitx}
\usepackage{longtable,tabularx}
\usepackage{bm}
\usepackage{overpic}
\usepackage{subcaption}
\usepackage{makecell}

\usepackage{scalerel}
\newcommand\wh[1]{\hstretch{.9}{\widehat{\hstretch{1.1111}{#1}}}}
\newcommand\wt[1]{\hstretch{.9}{\widetilde{\hstretch{1.1111}{#1}}}}

\usepackage[noend]{algpseudocode}
\usepackage{algorithmicx,algorithm}
\title{Rare-Event Chance-Constrained Flight Control Optimization Using Surrogate-Based Subset Simulation}

\author{Dalong Shi\footnote{Ph.D. Candidate, dalong.shi@tum.de.} and Florian Holzapfel\footnote{Professor, Associate Fellow AIAA, florian.holzapfel@tum.de.}}
\affil{Institute of Flight System Dynamics, Technical University of Munich, 85748 Garching, Germany}

\begin{document}

\maketitle

\begin{abstract}
A probabilistic performance-oriented control design optimization approach is introduced for flight systems. Aiming at estimating rare-event probabilities accurately and efficiently, subset simulation is combined with surrogate modeling techniques to improve efficiency. At each level of subset simulation, the samples that are close to the failure domain are employed to construct a surrogate model. The existing surrogate is then refined progressively. In return, seed and sample candidates are screened by the updated surrogate, thus saving a large number of calls to the true model and reducing the computational expense. Afterwards, control parameters are optimized under rare-event chance constraints to directly guarantee system performance. Simulations are conducted on an aircraft longitudinal model subject to parametric uncertainties to demonstrate the efficiency and accuracy of this method.
\end{abstract}

\section{Introduction}

\lettrine{V}{arious} uncertainties and disturbances are inevitable in real flight and might endanger the safety of aircraft. To guarantee flight safety, safety requirements of the flight control systems must be fulfilled with the existence of uncertainties and disturbances. Some safety requirements are defined as failure probability thresholds that the system must not exceed. For example, during automatic landing, the probabilities of exceeding safety limits must not be larger than the thresholds given in~\cite{csawo2003}. The acceptable failure probability thresholds are usually quite small (between $10^{-9}$ and $10^{-5}$), especially for failure events that may lead to severe safety issues~\cite{csawo2003, Loebl2015, Beaverstock2009}.

Probabilistic requirements are often formulated as chance (probabilistic) constraints, which have been frequently used in control community. The results in~\cite{Vitus2012} introduce both analytical methods and sampling-based methods to handle chance constraints, and the feedback controller is incorporated into the optimization with a risk allocation. A strategy based on split Bernstein polynomials and Markov chain Monte Carlo (MCMC) is implemented to estimate chance constraints for optimal control problems in~\cite{Zhao2015}. In~\cite{Piprek2019}, polynomial chaos expansion (PCE) and subset simulation (SuS) are employed to approximate rare-event probabilities within a chance-constrained open-loop optimal control framework. Besides, chance constraints are also common in model predictive control (MPC)~\cite{Mesbah2016}. Min-max theory is a very popular method for robust MPC and it tries to achieve a worst-case design to increase robustness~\cite{Diehl2004, Villanueva2017}. Chance constraints are converted into explicit algebraic constraints to assure the online applicability of MPC in~\cite{Gavilan2012}. In addition, PCE is utilized to propagate the parametric uncertainties and probabilistic constraints are transformed into the second-order cone constraints in~\cite{Mesbah2014}. Among these research, only the study~\cite{Piprek2019} is specifically tailored to rare failure probabilities, but the accuracy of the surrogate model in the rare failure domain may not be guaranteed.

SuS~\cite{Loebl2015, Au2001, Au2014} is an efficient method for estimating rare failure probabilities where the probability of a rare event is expressed as a product of much larger conditional probabilities. Although it achieves higher efficiency than Monte Carlo simulation (MCS), at least thousands of evaluations of the original model are required to reach a desirable accuracy~\cite{Li2016}. This is acceptable when estimating the probability of rare events, but for optimization problems with rare-event chance constraints, the computational cost is generally unaffordable.

To enhance the efficiency of SuS, many researchers combine it with surrogate models since they are usually analytical and it is very efficient to simulate with such  models. One possible scheme is to perform SuS using the surrogate model instead of the true model, but this necessitates a sufficiently accurate approximation of the true model. This goal has been successfully achieved by surrogate models such as kriging~\cite{Dubourg2011}, support vector machines (SVM)~\cite{Bourinet2011, Bourinet2016}, and neural networks (NN)~\cite{Papadopoulos2012}. In these works, adaptive training strategies that save a large number of calls to the true model are employed to construct highly accurate surrogates for the SuS process. An alternative scheme is to accelerate SuS with the surrogate model. For example, a delayed rejection strategy where the samples are first screened by the surrogate model is introduced during the MCMC sampling~\cite{Li2016}. Although the unbiasedness of this algorithm is guaranteed, the computational and statistical efficiency rely on the quality of the surrogate model.

In this paper, we combine SuS with surrogate models to accelerate the estimation of rare failure probabilities. This strategy is then applied to increase the efficiency of flight control optimization which directly ensures the satisfaction of rare-event chance constraints. In this approach, an initial global surrogate is first built using PCE. At each following SuS level, a local surrogate is constructed by adaptive response surface method (RSM) with samples close to the failure domain. The global surrogate is then updated by the local surrogate. Afterwards, in the subsequent SuS process, a substantial proportion of calls to the true model are substituted by the calls to the refined surrogate.

The rest of the paper is organized as follows. Section~\ref{sec:RBCD} presents the framework of performance-guaranteed control optimization with chance constraints. Section~\ref{sec:basics} recalls the basics of SuS, PCE, and RSM. Section~\ref{sec:SBSS} proposes the surrogate-based subset simulation (SBSS) method to progressively refine the surrogate model and accelerate the SuS procedure. In Section~\ref{sec:eg}, the introduced framework and method are implemented on a flight control optimization problem with rare-event probabilistic requirements.

\section{Performance-Guaranteed Control Optimization}
\label{sec:RBCD}

Consider a class of closed-loop dynamic systems subject to parametric uncertainties:
\begin{equation} \label{eq:sys}
\left\{
\begin{array}{ll}
\dot{\bm{x}}(t) = \bm{f}(\bm{x}(t), \bm{u}(t), \bm{\theta}, \bm{k}), \\
\bm{y}(t) = \bm{g}(\bm{x}(t), \bm{u}(t), \bm{\theta}), \\
\end{array}
\right.
\end{equation}
where $\bm{x} \in {\mathbb{R}^{n_x}}$ represents the state vector, $\bm{u} \in {\mathbb{R}^{n_u}}$ represents the input vector, and $\bm{y} \in {\mathbb{R}^{n_y}}$ represents the output vector. Uncertain parameters are denoted by $\bm{\theta} \in {\mathbb{R}^{p}}$, and $\bm{k} \in {\mathbb{R}^{d}}$ contains control gains that are uncorrelated with $\bm{\theta}$. The functions $\bm{f}(\cdot)$ and $\bm{g}(\cdot)$ describe the system dynamics and are potentially unknown. The performance of the dynamic system can be represented by a series of performance functions $h_i(\bm{\theta}, \bm{k}), i = 0, 1, \ldots, n_h$.

The control optimization that guarantees system performance is stated as follows:
\begin{equation} \label{eq:opt}
\begin{array}{llll}
&\min \limits_{\bm{k}}\quad & \mathop{\mathbb{P}}[h_0(\bm{\theta}, \bm{k}) < C_0], \\
&{\rm s.t.} &\mathop{\mathbb{P}}[h_i(\bm{\theta}, \bm{k}) < C_i] < \beta_i, \quad & i = 1, \ldots, n_h, \\
&           &c_i(\bm{k}) \leq 0,  & i = 1, \ldots, n_c,
\end{array}
\end{equation}
where $C_i, i = 0, 1, \ldots, n_h,$ are the limits for each performance function, $\beta_i, i = 1, \ldots, n_h,$ are the probabilistic thresholds that should not be violated, and $c_i(\bm{k})\leq 0, i = 1, \ldots, n_c,$ are deterministic constraints. This control optimization framework aims at minimizing the exceeding probability of $h_0(\bm{\theta}, \bm{k})$ while ensuring that other exceeding probabilities of performance functions are within their corresponding safe ranges. Consequently, it is able to directly fulfill the statistical requirements and explore further performance.

\section{Basics on Subset Simulation and Surrogate Model Construction}
\label{sec:basics}

\subsection{Subset Simulation}

The failure probability can be generally expressed as follows:
\begin{equation} \label{eq:pf}
	p_f = \int_F\rho(\bm{\theta}) \mathrm{d} \bm{\theta},
\end{equation}
where $\rho(\bm{\theta})$ is the joint probability density function (PDF) of $\bm{\theta}$, and $F = \{ \bm{\theta}\in \mathbb{R}^{p} : h(\bm{\theta}) \leq 0 \}$ is the failure domain, in which $h(\bm{\theta})$ is the performance function with fixed control gains.

The key idea of SuS is to introduce $m$ intermediate failure domains $F_j, j = 1, \ldots, m,$ satisfying $F_1 \supset F_2 \supset \cdots \supset F_m = F$, such that the failure probability of a rare event is transcribed into a product of conditional probabilities~\cite{Au2001, Au2014}:
\begin{equation} \label{eq:ss}
	p_f = \mathop{\mathbb{P}}[F_m] = \mathop{\mathbb{P}}[F_m|F_{m-1}] \mathop{\mathbb{P}}[F_{m-1}] = \cdots = \prod_{j=1}^{m} \mathop{\mathbb{P}}[F_{j}|F_{j-1}],
\end{equation}
where $\mathop{\mathbb{P}}[\cdot]$ is the probability operator, $F_0$ is the uncertain parameter space, and $\mathop{\mathbb{P}}[F_{1}|F_{0}] = \mathop{\mathbb{P}}[F_{1}]$. In practice, intermediate failure domains are defined as $F_j = \{ \bm{\theta}\in \mathbb{R}^{p} : h(\bm{\theta}) \leq b_j \}$ with $b_1 > b_2 > \cdots > b_m = 0$, where $b_j, j = 1, \ldots, m,$ are intermediate thresholds. The thresholds $b_j, j = 1, \ldots, m-1,$ are selected reasonably such that the conditional probabilities $p_j = \mathop{\mathbb{P}}[F_j|F_{j-1}], j = 1, \ldots, m,$ are large enough ($p_j \in [0.1, 0.3]$ is suggested for the best performance~\cite{Zuev2012}) to be efficiently estimated by simulation.

The basic SuS algorithm can be summarized as given in Algorithm~\ref{alg:ss}. Various MCMC sampling strategies have been developed to enhance the performance of SuS, such as the modified (component-wise) Metropolis-Hastings algorithm~\cite{Au2001}, the modified Metropolis-Hastings algorithm with delayed rejection~\cite{Zuev2011}, and the Gaussian conditional sampling~\cite{Papaioannou2015}. In this paper, the Gaussian conditional sampling is applied, in which the candidate samples generated from the proposal PDF always differ from the current sample. Therefore, the number of samples generated at each subset level ($j = 1, \ldots, m-1$) is
\begin{equation} \label{eq:ss1}
(1-p_0)N,
\end{equation}
and the total number of calls to the true model is
\begin{equation} \label{eq:ss2}
N+(m-1)(1-p_0)N.
\end{equation}

\begin{algorithm} [t!]
	\caption{Basic SuS algorithm \cite{Au2001, Au2014}}
	\label{alg:ss} 
	\begin{algorithmic}[1]
		\Require Performance function $h(\bm{\theta})$; The distribution of uncertain parameters $\rho(\bm{\theta})$; The number of samples at each level $N$; Conditional probability $p_0$.
		\Ensure The estimate of failure probability $\wh{p}_f$.
		\State Generate $N$ independent and identically distributed (i.i.d.) samples $\{\bm{\theta}_0^{(i)}: i = 1, \ldots, N \}$ according to $\rho(\bm{\theta})$;
		\State Calculate the corresponding function values $\{ h(\bm{\theta}_0^{(i)}): i = 1, \ldots, N \}$;
		\State Sort these function values in ascending order, find the $p_0$-percentile $b_1$, and set $F_1 = \{ \bm{\theta}\in \mathbb{R}^{p}: h(\bm{\theta}) \leq b_1 \}$;
		\State Set $j = 1$;
		\While{$b_j > 0$}
		\State Choose samples $\bm{\theta}_{j-1}^{(i)} \in F_j$ as seeds $\{ \bm{\theta}_{j-1, \mathrm{seed}}^{(i)}: i = 1, \ldots, N_s \}$, where $N_s = p_0 N$ is an integer;
		\State Generate $N$ samples $\{ \bm{\theta}_j^{(i)}: i = 1, \ldots, N \}$ from the seeds using MCMC sampling;
		\State Calculate function values $\{ h(\bm{\theta}_j^{(i)}): i = 1, \ldots, N \}$;
		\State Sort them in ascending order, find the $p_0$-percentile $b_{j+1}$, and set $F_{j+1} = \{ \bm{\theta}\in \mathbb{R}^{p} : h(\bm{\theta}) \leq b_{j+1} \}$;
		\State Set $j = j+1$;
		\EndWhile
		\State {\bf end while}
		\State Obtain the number of subsets $m = j$;
		\State Identify the number of samples $\bm{\theta}_{m-1}^{(i)} \in F$: $n_f$;
		\State Estimate the failure probability $\wh{p}_f = p_0^{m-1} \frac{n_f}{N}$;
		\State \Return $\wh{p}_f$.
	\end{algorithmic}
\end{algorithm}

The coefficient of variation (c.o.v.) of the estimated failure probability can be evaluated by the c.o.v. of the estimated intermediate conditional probabilities, see~\cite{Au2001} for details.

\subsection{Polynomial Chaos Expansion}

The PCE of $h(\bm{\theta})$ can be represented as an infinite weighted sum of polynomial bases~\cite{Xiu2002, Xiu2010}:
\begin{equation} \label{eq:PCE0}
	h(\bm{\xi}) = \sum_{i=0}^{\infty}a_i\Psi_i(\bm{\xi}),
\end{equation}
where $\bm{\xi} \in {\mathbb{R}^{p}}$ denotes a vector of standard random variables, $\Psi_i(\bm{\xi})$ denote the multivariate polynomial basis functions, and $a_i$ denote the corresponding expansion coefficients. In practical problems, $\bm{\theta}$ is usually not a vector of standard random variables, therefore it is necessary to  transform $\bm{\theta}$ into a set of standard variables $\bm{\xi}$ through the isoprobabilistic transformation:
\begin{equation} \label{eq:iso}
	\bm{\xi} = \tau(\bm{\theta}).
\end{equation}
The multivariate polynomials $\Psi_i(\bm{\xi})$ can be constructed as the product of their univariate counterparts:
\begin{equation} \label{eq:PCE1}
	\Psi_i(\bm{\xi}) = \prod_{r=1}^{p}\psi_{m_r^i}(\xi_{r}),
\end{equation}
where $m_r^i$ denote the multi-indexes that contain all the possible combinations of univariate polynomials, and $\psi_{m_r^i}(\xi_{r})$ denote the $m_r^i$th-degree univariate orthogonal polynomial bases that satisfy
\begin{equation} \label{eq:PCE2}
	\mathop{\mathbb{E}}[\psi_i(\xi) \psi_j(\xi)] = \int_\Omega \psi_i(\xi) \psi_j(\xi) \rho(\xi) \mathrm{d} \xi = \gamma_i \delta_{ij}, \quad i, j \in \mathop{\mathbb{N}},
\end{equation}
where $\mathop{\mathbb{E}}[\cdot]$ is the expectation with respect to $\rho(\xi)$, $\Omega$ is the support of $\xi$, $\gamma_i = \mathop{\mathbb{E}} \left[ \psi_i^2(\xi) \right]$, and $\delta_{i j}$ is the Kronecker function equal to $1$ when $i = j$ and $0$ otherwise. Consequently, the orthogonality also holds for $\Psi_i(\bm{\xi})$:
\begin{equation} \label{eq:PCE3}
	\mathop{\mathbb{E}}[\Psi_i(\bm{\xi}) \Psi_j(\bm{\xi})] = \gamma_i \delta_{ij}, \quad i, j \in \mathop{\mathbb{N}},
\end{equation}
where $\gamma_i = \mathop{\mathbb{E}} \left[ \Psi_i^2(\bm{\xi}) \right]$.

The polynomial bases $\psi_i$ can be selected from Table~\ref{table:orth} based on the distribution of $\xi$. Table~\ref{table:orth} shows classical families of orthogonal polynomials in the Wiener-Askey scheme.

\begin{table}[t!]
	\centering
	\caption{Classical families of orthogonal polynomials \cite{Xiu2002}.}\label{table:orth}
	\begin{tabular}{cccc}
		\hline
		Type of variable $\xi$ & Distribution $\rho(\xi)$ & Support $\Omega$ & Orthogonal polynomial $\psi_n(\xi)$ \\\hline
		Gaussian & $\frac{1}{\sqrt{2\pi}} e^{-\xi^{2}/2}$ & $(-\infty, \infty)$ & Hermite $H_n(\xi)$ \\
		Uniform & $\frac{1}{2}$ & $[-1,1]$ & Legendre $P_n(\xi)$ \\
		Gamma & $\frac{\xi^{\alpha} e^{-\xi}}{\Gamma(\alpha+1)}$ & $[0, \infty)$ & Laguerre $L_n^{\alpha}(\xi)$ \\ 
		Beta & $\frac{(1-\xi)^{\alpha}(1+\xi)^{\beta}}{2^{\alpha+\beta+1} B(\alpha+1, \beta+1)}$ & $[-1,1]$ & Jacobi $J_n^{\alpha,\beta}(\xi)$ \\ \hline
	\end{tabular}
\end{table}

The expansion with infinite terms in Eq.~\eqref{eq:PCE0} is exact, but in practice, only a finite number of terms can be computed. Usually, the expansion is truncated up to a certain degree $M$ and the performance function is approximated by the surrogate model $\wh{h}$ built with the $M$-th order PCE:
\begin{equation} \label{eq:PCE4}
	h(\bm{\xi}) \approx \wh{h}(\bm{\xi}) = \sum_{i=0}^{P-1}a_i\Psi_i(\bm{\xi}),
\end{equation}
where the number of terms is generally given by
\begin{equation} \label{eq:PCE5}
	P = \frac{(M+p)!}{M!p!}.
\end{equation}

Several non-intrusive methods have been developed to compute the expansion coefficients, among which least squares regression and spectral projection are the most fundamental and widely used techniques.

\subsubsection{Least squares regression}

Least squares regression aims at finding a set of coefficients $\bm{a} = [a_0, a_1, \ldots, a_{P-1}]^\mathsf{T}$ that minimize the mean square residual error with a collection of samples (experimental design) $\{\bm{\xi}^{(i)}: i = 1, 2, \ldots, n_s\}$ and the associated model evaluations $\bm{y} = [h(\bm{\xi}^{(1)}), h(\bm{\xi}^{(2)}), \ldots, h(\bm{\xi}^{(n_s)})] ^\mathsf{T}$~\cite{Sudret2014}: 
\begin{equation} \label{eq:lsr1}
\bm{a} = \arg \min \mathop{\mathbb{E}} \left[\left( h(\bm{\xi})-\wh{h}(\bm{\xi}) \right)^2 \right] = \left( \bm{A}^\mathsf{T} \bm{A} \right)^{-1} \bm{A}^\mathsf{T} \bm{y},
\end{equation}
where $\bm{A}$ is the experimental matrix with elements
\begin{equation} \label{eq:lsr2}
\bm{A}_{ij} = \Psi_{j-1}(\bm{\xi}^{(i)}), \quad i = 1, \ldots, n_s; j = 1, \ldots, P.
\end{equation}
A rule of thumb for designating the size of experimental design is $n_s \approx 2P$--$3P$~\cite{Sudret2014}.

\subsubsection{Spectral projection}

Due to the orthogonality of polynomial bases, one can project the random response in Eq.~\eqref{eq:PCE4} against each basis and obtain the expansion coefficients~\cite{Xiu2007}:
\begin{equation} \label{eq:projection1}
	a_i = \frac{1}{\gamma_i} \mathop{\mathbb{E}} [h(\bm{\xi})\Psi_i(\bm{\xi})] = \frac{1}{\gamma_i} \int_{\bm{\Omega}} h(\bm{\xi})\Psi_i(\bm{\xi})\rho(\bm{\xi})\, \mathrm{d} \bm{\xi}.
\end{equation}
As a result, the residual error $\epsilon_r = h(\bm{\xi})-\wh{h}(\bm{\xi})$ is orthogonal to the chosen bases. The integral in Eq.~\eqref{eq:projection1} can be evaluated numerically by Gaussian quadrature or sparse quadrature, which is a weighted-sum scheme:
\begin{equation} \label{eq:projection2}
	a_i \approx \frac{1}{\gamma_i} \sum_{j=1}^{Q}h(\bm{\xi}^{(j)})\Psi_i(\bm{\xi}^{(j)})w^{(j)},
\end{equation}
where $\bm{\xi}^{(j)}$ are the quadrature nodes, $ w^{(j)}$ are the corresponding weights, and $Q$ is the number of these quadrature nodes.

\subsection{Response Surface Method}

RSM~\cite{Khuri2010} is a useful tool to establish the relationship between input variables and model responses. Generally, the relationship, i.e., response surface or surrogate model, is assumed to be a low-order polynomial with unknown coefficients and the coefficients can be obtained by least squares regression. This is similar to PCE using regression method, and the latter can be regarded as an extension of the former to the standard random space. In RSM, the orthogonality as shown in Eq.~\eqref{eq:PCE3} is not fulfilled, but one can still employ the framework in Eq.~\eqref{eq:lsr1} to achieve the response surface construction.

\section{Surrogate-Based Subset Simulation}
\label{sec:SBSS}

\subsection{Subset Simulation Based on Response Surface Method}

Even though SuS is very efficient in estimating rare-event probabilities, at least thousands of calls to the true model are still required to achieve a sufficient accuracy. To alleviate this burden, we propose the SBSS method to refine the surrogate model of the performance function $h(\bm{\theta})$ in the domain of interest and accelerate the conventional SuS with the updated surrogate model.

At the initial level of SuS where the crude MCS is performed, a global surrogate model $\wh{h}_0(\bm{\theta})$ is constructed by PCE. This surrogate model is then used to select seeds from the initial $N$ samples. Considering that the potential error of the surrogate, we do not determine the $N_s = p_0N$ seeds directly. Instead, we choose $\widetilde{N}_s = \widetilde{p}_0N$ seed candidates which are closest to the failure domain in the viewpoint of the surrogate model. Afterwards, the $\widetilde{N}_s$ candidates are evaluated with the true model and $N_s$ seeds are selected from these candidates. Then, we build a local surrogate model using RSM with the $\widetilde{N}_s$ samples, and refine the existing global surrogate. After that, the updated surrogate is employed in MCMC sampling to decide whether to accept or reject the generated samples. This means that additional calls to the true model are not required in MCMC sampling. The steps to choose seeds, refine the surrogate model, and accept or reject samples in MCMC sampling are repeated until the end of the SuS procedure.

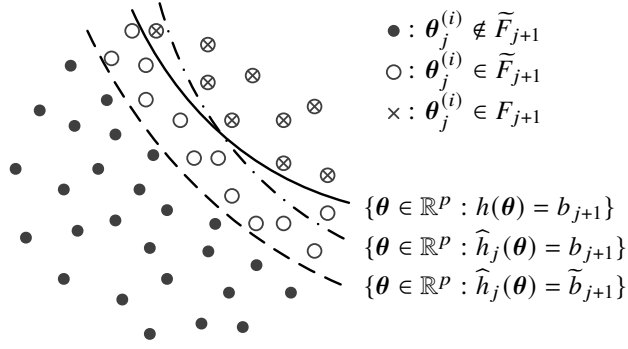
\begin{figure}[t!]
	\centering
	\begin{overpic}[width=8.5cm]{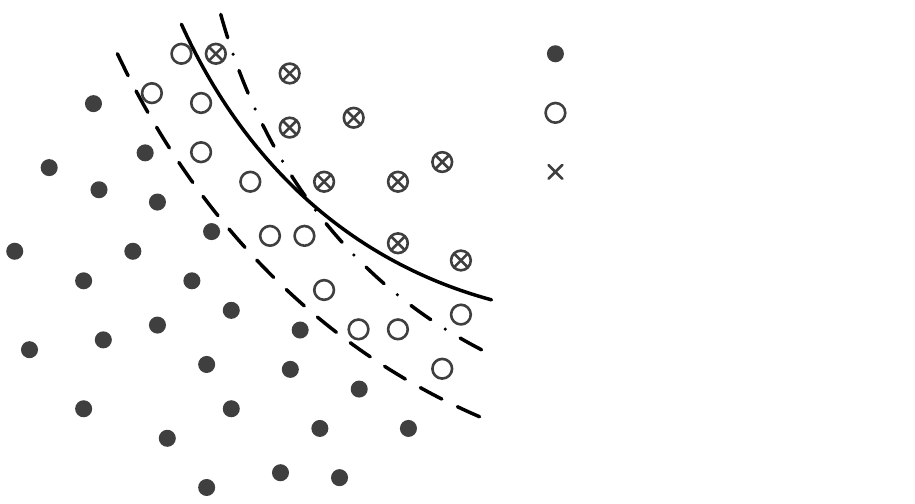}
		\put(56, 20.4){$\{ \bm{\theta}\in \mathbb{R}^{p}:{h}(\bm{\theta})={b}_{j+1} \}$}
		\put(56, 14.2){$\{ \bm{\theta}\in \mathbb{R}^{p}:\wh{h}_j(\bm{\theta}) = {b}_{j+1} \}$}
		\put(56, 8){$\{ \bm{\theta}\in \mathbb{R}^{p}:\wh{h}_j(\bm{\theta}) = \wt{b}_{j+1} \}$}
		\put(63, 48.30){: $\bm{\theta}_j^{(i)} \notin \widetilde{F}_{j+1}$}
		\put(63, 41.87) {: $\bm{\theta}_j^{(i)} \in \widetilde{F}_{j+1}$}
		\put(63, 35.44){: $\bm{\theta}_j^{(i)} \in F_{j+1}$}
	\end{overpic}
	\caption{Samples at each level of SBSS.}
	\label{fig:samples}
\end{figure}

At each subset level ($j = 0, 1, \ldots, {m-1}$), we choose $\widetilde{N}_s$ seed candidates from the $N$ samples using the updated surrogate model. In practice, we achieve this by determining an intermediate threshold $\wt{b}_{j+1}$ such that $\widetilde{N}_s$ samples are in the intermediate failure domain $\widetilde{F}_{j+1} = \{ \bm{\theta}\in \mathbb{R}^{p}: \wh{h}_j (\bm{\theta}) \leq \wt{b}_{j+1} \}$, where $\wh{h}_j (\bm{\theta})$ represents the refined surrogate model at $j$-th level. Similarly, an intermediate threshold $b_{j+1}$ is determined so that $N_s$ seeds are in the intermediate failure domain $F_{j+1} = \{ \bm{\theta}\in \mathbb{R}^{p}: h(\bm{\theta}) \leq b_{j+1} \}$. These different intermediate thresholds and domains are depicted in Fig.~\ref{fig:samples}. As stated before, the margin between $\{ \bm{\theta}\in \mathbb{R}^{p}: \wh{h}_j (\bm{\theta}) = \wt{b}_{j+1} \}$ and $\{ \bm{\theta}\in \mathbb{R}^{p}: \wh{h}_j (\bm{\theta}) = {b}_{j+1} \}$ allows limited error of the surrogate model.

The global surrogate model is refined progressively as follows:
\begin{equation} \label{eq:sbss1}
\wh{h}_{j+1}(\bm{\theta}) = \left\{
\begin{array}{ll}
\wh{h}_{j}(\bm{\theta}), &\wh{h}_{j}(\bm{\theta}) > \wt{b}_{j+1};\\
\wh{h}_{j+1, \rm{local}}(\bm{\theta}), \quad &\wh{h}_{j}(\bm{\theta}) \leq \wt{b}_{j+1},\\
\end{array}
\right.
\end{equation}
where $\wh{h}_{j+1, \rm{local}}(\bm{\theta})$ denotes the local surrogate model constructed by RSM at $j$-th level. The global surrogate is expressed as a piecewise function with jump discontinuities between subdomains. The influence of the discontinuities can be alleviated by selecting a larger $\wt{p}_{0}$, but this increases the number of calls to the true model. It is thereby a tradeoff between accuracy and computational expense.

In summary, the proposed strategy refines the global surrogate model gradually as the samples approach the failure domain, and in return a significant number of true model evaluations are approximated by calling the surrogate model. A detailed algorithm is given in Algorithm~\ref{alg:sbss}. It is beneficial, especially when the evaluation of the true model is computationally demanding. The number of calls to the true model at each subset level is
\begin{equation} \label{eq:sbss2}
\wt{p}_0N,
\end{equation}
and the total number of calls is
\begin{equation} \label{eq:sbss3}
N_0 + (m-1)\wt{p}_0N,
\end{equation}
where $N_0$ is the number of true model evaluations when building the initial surrogate $\wh{h}_0 (\bm{\theta})$ using PCE.

\begin{algorithm}[t!]
	\caption{Basic SBSS algorithm}
	\label{alg:sbss}
	\begin{algorithmic}[1]
		\Require Performance function $h(\bm{\theta})$; The distribution of uncertain parameters $\rho(\bm{\theta})$; The number of samples at each level $N$; Conditional probability $p_0$; The percentage of samples for the construction of local surrogates $\wt{p}_0$ $(\wt{p}_0 \geq p_0)$.
		\Ensure The estimate of failure probability $\wh{p}_f$.
		\State Generate $N$ i.i.d. samples $\{ \bm{\theta}_0^{(i)}: i = 1, \ldots, N \}$ according to $\rho(\bm{\theta})$ and construct a global surrogate model $\wh{h}_0 (\bm{\theta})$ using PCE;
		\State Calculate the function values $\{ \wh{h}_0(\bm{\theta}_0^{(i)}): i = 1, \ldots, N \}$;
		\State Sort these function values in ascending order, find the $\wt{p}_0$-percentile $\wt{b}_1$, and set $\widetilde{F}_1 = \{ \bm{\theta}\in \mathbb{R}^{p}: \wh{h}_0 (\bm{\theta}) \leq \wt{b}_1 \}$;
		\State Select samples $\bm{\theta}_{0}^{(i)} \in \widetilde{F}_1$ as the experimental design $\{ \bm{\theta}_{0,\rm{ED}}^{(i)}: i = 1, \ldots, \widetilde{N}_s \}$, where $\widetilde{N}_s = \wt{p}_0N$;
		\State Calculate the function values $\{ h(\bm{\theta}_{0,\rm{ED}}^{(i)}): i = 1, \ldots, \widetilde{N}_s \}$;
		\State Sort these values in ascending order, find the $p_0$-percentile $b_1$, and set $F_1 = \{ \bm{\theta}\in \mathbb{R}^{p}: h(\bm{\theta}) \leq b_1 \}$;
		\State Set $j = 1$;
		\While{$b_j > 0$}
		\State \parbox[t]{\dimexpr\linewidth-\algorithmicindent}
		{Build a local surrogate $\wh{h}_{j,\rm{local}} (\bm{\theta})$ implementing RSM with the experimental design $\{ \bm{\theta}_{j-1,\rm{ED}}^{(i)}: i = 1, \ldots, \widetilde{N}_s \}$ and the corresponding responses $\{ h(\bm{\theta}_{j-1,\rm{ED}}^{(i)}): i = 1, \ldots, \widetilde{N}_s \}$;}
		\State Refine the global surrogate model as in Eq.~\eqref{eq:sbss1};
		\State Choose samples $\bm{\theta}_{j-1}^{(i)} \in F_j$ as seeds $\{ \bm{\theta}_{j-1, \mathrm{seed}}^{(i)}: i = 1, \ldots, N_s \}$, where $N_s = p_0 N$;
		\State \parbox[t]{\dimexpr\linewidth-\algorithmicindent}
		{Generate $N$ samples $\{ \bm{\theta}_j^{(i)}: i = 1, \ldots, N \}$ from the seeds applying MCMC sampling, wherein samples are accepted or rejected using the updated global surrogate model $\wh{h}_j (\bm{\theta})$;}
		\State Calculate the function values $\{ \wh{h}_j(\bm{\theta}_j^{(i)}): i = 1, \ldots, N \}$;
		\State \parbox[t]{\dimexpr\linewidth-\algorithmicindent}
		{Sort these values in ascending order, find the $\wt{p}_0$-percentile $\wt{b}_{j+1}$, and set $\widetilde{F}_{j+1} = \{ \bm{\theta}\in \mathbb{R}^{p}: \wh{h}_j (\bm{\theta}) \leq \wt{b}_{j+1} \}$;}
		\State Select samples $\bm{\theta}_{j}^{(i)} \in \widetilde{F}_{j+1}$ as the experimental design $\{ \bm{\theta}_{j,\rm{ED}}^{(i)}: i = 1, \ldots, \widetilde{N}_s \}$;
		\State Calculate the function values $\{ h(\bm{\theta}_{j,\rm{ED}}^{(i)}): i = 1, \ldots, \widetilde{N}_s \}$;
		\State \parbox[t]{\dimexpr\linewidth-\algorithmicindent}
		{Sort these values in ascending order, find the $p_0$-percentile $b_{j+1}$, and set $F_{j+1} = \{ \bm{\theta}\in \mathbb{R}^{p}: h(\bm{\theta}) \leq b_{j+1} \}$;}
		\State Set $j = j+1$;
		\EndWhile
		\State {\bf end while}
		\State Obtain the number of subsets $m = j$;
		\State Identify the number of samples $\bm{\theta}_{m-1}^{(i)} \in F$: $n_f$;
		\State Estimate the failure probability $\wh{p}_f = p_0^{m-1} \frac{n_f}{N}$;
		\State \Return $\wh{p}_f$.
	\end{algorithmic}
\end{algorithm}

\subsection{Adaptive Response Surface Method}

The quality of the constructed surrogate model can be measured by the mean square error of the residual (i.e., empirical error):
\begin{equation} \label{err_emp}
	\epsilon_{\rm{emp}} = \frac{1}{n_s} \sum_{i=1}^{n_s} \left( h(\bm{\theta}^{(i)})-\wh{h}(\bm{\theta}^{(i)}) \right)^2. 
\end{equation}
Its normalized quantity, which is called relative empirical error, is computed by
\begin{equation} \label{err_remp}
	\epsilon_{\rm{emp}}^{r} = \frac{\epsilon_{\rm{emp}}}{{\rm var}[\bm{y}]},
\end{equation}
where ${\rm var}[\bm{y}]$ is the variance of responses $\bm{y}$. The error is reduced with the increase of the order of RSM until the response surface fits the samples in experimental design perfectly, i.e., $\epsilon_{\rm emp}$ or $\epsilon_{\rm emp}^r$ is almost zero. However, with sufficiently high order, the risk involved is that the approximation of these samples can be extremely good but very bad elsewhere. This situation is known as overfitting, which means the response surface would be quite different with that built by another set of samples. Therefore, the surrogate modeling error is usually underestimated by the empirical error, but can be better estimated by the leave-one-out (LOO) cross-validation error \cite{Sudret2014}. LOO cross-validation removes one point $\bm{\theta}^{(i)}$ of the experimental design and construct a surrogate model denoted by $\wh{h}^{(-i)}(\bm{\theta})$ from the remaining $n_s-1$ samples. The predicted residual error at $\bm{\theta}^{(i)}$ reads \cite{Sudret2014}:
\begin{equation} \label{err_loo0}
	\epsilon_{i} = h(\bm{\theta}^{(i)}) - \wh{h}^{(-i)}(\bm{\theta}^{(i)}) = \frac{h(\bm{\theta}^{(i)})-\wh{h}(\bm{\theta}^{(i)})}{1-s_i},
\end{equation}
where $s_i$ is the $i$-th diagonal term of matrix $\bm{A} \left( \bm{A}^\mathsf{T} \bm{A} \right)^{-1} \bm{A}^\mathsf{T}$. The LOO error is defined as
\begin{equation} \label{err_loo}
	\epsilon_{\rm{LOO}} = \frac{1}{n_s} \sum_{i=1}^{n_s} \epsilon_{i}^2 = \frac{1}{n_s} \sum_{i=1}^{n_s} \left( \frac{h(\bm{\theta}^{(i)})-\wh{h}(\bm{\theta}^{(i)})}{1-s_i} \right)^2. 
\end{equation}
Similarly, the relative LOO error can be obtained by
\begin{equation} \label{err_rloo}
\epsilon_{\rm LOO}^{r} = \frac{\epsilon_{\rm LOO}}{{\rm var}[\bm{y}]}.
\end{equation}

The adaptive RSM that minimizes the LOO error is summarized in Algorithm~\ref{alg:arsm}. By this means, excessively high orders are prohibited and the overfitting problem is avoided.

\begin{algorithm} [t!]
	\caption{Adaptive RSM algorithm}
	\label{alg:arsm}
	\begin{algorithmic}[1]
		\Require Performance function $h(\bm{\theta})$; The number of uncertain parameters $p$; The possible orders of the response surface $M_{\rm min}: M_{\rm max}$; The size of experimental design $n_s$.
		\Ensure The optimal surrogate model $\wh{h}(\bm{\theta})$
		\State Obtain the experimental design $\{ \bm{\theta}^{(i)}: i = 1, \ldots, n_s \}$;
		\State Calculate the corresponding function values $\{ h(\bm{\theta}^{(i)}): i = 1, \ldots, n_s \}$;
		\For {$M = M_{\rm min}: M_{\rm max}$}
		\State Generate polynomial bases $\{ \Psi_i(\bm{\theta}): i = 0, 1, \ldots, P-1 \}$;
		\State Calculate the experimental matrix $\bm{A}$ as in Eq.~\eqref{eq:lsr2};
		\State Solve the least squares problem in Eq.~\eqref{eq:lsr1};
		\State Compute $\epsilon_{\rm LOO}^r(M)$ according to Eqs.~\eqref{err_loo} and \eqref{err_rloo};
		\EndFor
		\State {\bf end for}
		\State $M^* = \arg \min \epsilon_{\rm LOO}^r(M)$;
		\State Obtain the optimal surrogate model $\wh{h}(\bm{\theta})$;
		\State \Return $\wh{h}(\bm{\theta})$.
	\end{algorithmic}
\end{algorithm}

\section{Application to Flight Control}
\label{sec:eg}

\subsection{Simulation Model}

The control plant is an aircraft (Diamond DA42) longitudinal model integrating actuator dynamics, structural mode and filters. A proportional-integral-derivative (PID) controller with feedforward control is implemented:
\begin{equation}
	\dot{q}_{\rm cmd} = k_H n_{z,\rm cmd} + k_{n_z} n_{z} + k_I \int \left(n_{z,\rm cmd} - n_{z} \right) \mathrm{d} t + k_q \omega_y,
\end{equation}
where $\dot{q}_{\rm cmd}$ is the pitch acceleration command, $\omega_y$ is the feedback signal of pitch angular rate, and $n_{z}$ and $n_{z,\rm cmd}$ are the feedback signal of vertical load factor and its command. $\bm{k} = [k_H, k_{n_z}, k_I, k_q]^\mathsf{T}$ is the vector of control gains.

Uncertainties in aerodynamic derivatives are considered and the relative values with regard to the reference values are assumed to be Gaussian-distributed:
\begin{equation}
\left[ \begin{array}{c} M_{\alpha}/{M_{\alpha,\rm ref}} \\ M_q/{M_{q,\rm ref}} \\ M_{\eta}/{M_{\eta,\rm ref}} \end{array} \right] 
\sim \mathcal N \left( \left[ \begin{array}{c} \mu_{\alpha} \\ \mu_q \\ \mu_{\eta} \end{array} \right],
\left[ \begin{array}{ccc} \sigma_{\alpha}^2 & 0 & 0 \\ 0 & \sigma_q^2 & 0 \\ 0 & 0 & \sigma_{\eta}^2 \end{array} \right] \right), 
\end{equation}
where $M_{\alpha}$, $M_q$, and $M_{\eta}$ are the aerodynamic moments about the angle of attack $\alpha$, the pitch rate $q$, and the elevator deflection $\eta$, respectively. The mean $\mu_{\alpha} = \mu_q = \mu_{\eta} = 1$ and the standard deviation $\sigma_{\alpha} = \sigma_q = \sigma_{\eta} = 0.15$.

In the end, the closed-loop system is linear and can be represented as
\begin{equation}
	\left\{
	\begin{array}{ll}
	\dot{\bm{x}} = \bm{A}_{c} \bm{x} + \bm{B}_{c} \bm{u},\\
	\bm{y} = \bm{C}_{c} \bm{x},
	\end{array}
	\right.
\end{equation}
where $\bm{A}_{c}$, $\bm{B}_{c}$, and $\bm{C}_{c}$ are closed-loop system matrices. The dependence of these system matrices on $\bm{k}$ and uncertain aerodynamic derivatives has been omitted to simplify the notation. Besides the state introduced by the integration element of the control structure, the state vector $\bm{x}$ contains the states of short period dynamics ($\alpha$ and $q$), actuator dynamics, structural mode, and notch filter. Each of them is modeled as a second-order system. The output $\bm{y} = n_z$ and the input vector $\bm{u} = [n_{z,\rm cmd}, w_z]^\mathsf{T}$, where $w_z$ is the vertical wind velocity of the standard ``$1-$cosine'' gust \cite{Moorhouse1980}:
\begin{equation}
	w_z = \left\{
	\begin{array}{ll}
	\hspace{0.06cm} 0, & x_g <0; \\
	\dfrac{v_g}{2} \left[1-\cos\left(\dfrac{\pi x_g}{d_g} \right) \right], \quad & 0 \leq x_g < d_g; \\
	\hspace{0.03cm} v_g, & x_g \geq d_g, \\
	\end{array}
	\right.
\end{equation}
where $x_g$ is the traveled horizontal distance, $d_g$ is the gust length, and $v_g$ is the gust amplitude. In this application, $d_g=91.4~\mathrm{m}$ and $v_g=13.9~\mathrm{m/s}$.

\subsection{Rare-Event Probability Estimation}

Consider the stability margin requirements \cite{SAE2007} as follows:
\begin{equation} \label{eq:sm}
\begin{array}{ll}
\mathop{\mathbb{P}}[h_{1}(\bm{\theta}) < 6~\rm dB] < 10^{-6}, \\
\mathop{\mathbb{P}}[h_{2}(\bm{\theta}) < 45 ^{\circ}] < 10^{-6},
\end{array}
\end{equation}
where $h_{1}(\bm{\theta})$ and $h_{2}(\bm{\theta})$ denote the performance functions of gain margin (GM) and phase margin (PM), respectively, given fixed control gains. Note that no analytical form is available for these performance functions. To directly guarantee the compliance with the rare-event constraints, the rare probabilities must be precisely estimated first. In the following part, the proposed SBSS method is applied to accomplish this task.
The initial surrogate model is constructed by the 5th-order PCE (spectral projection) and a tensor product quadrature with 6 nodes in each dimension (216 nodes in total). The Gaussian conditional sampling \cite{Papaioannou2015} is implemented in both SuS and SBSS, and other settings are $N = 2000$, $p_0 = 0.1$, $\wh{p}_0 = 0.11$, and $M_{\rm min}:M_{\rm max} = 2:7$.

\begin{figure}[t!]
	\centering
	\subcaptionbox{The relative errors of $\wh{h}_1(\bm{\theta})$ \label{fig:lsr_err_gm}}
	{\includegraphics[width=.460\textwidth]{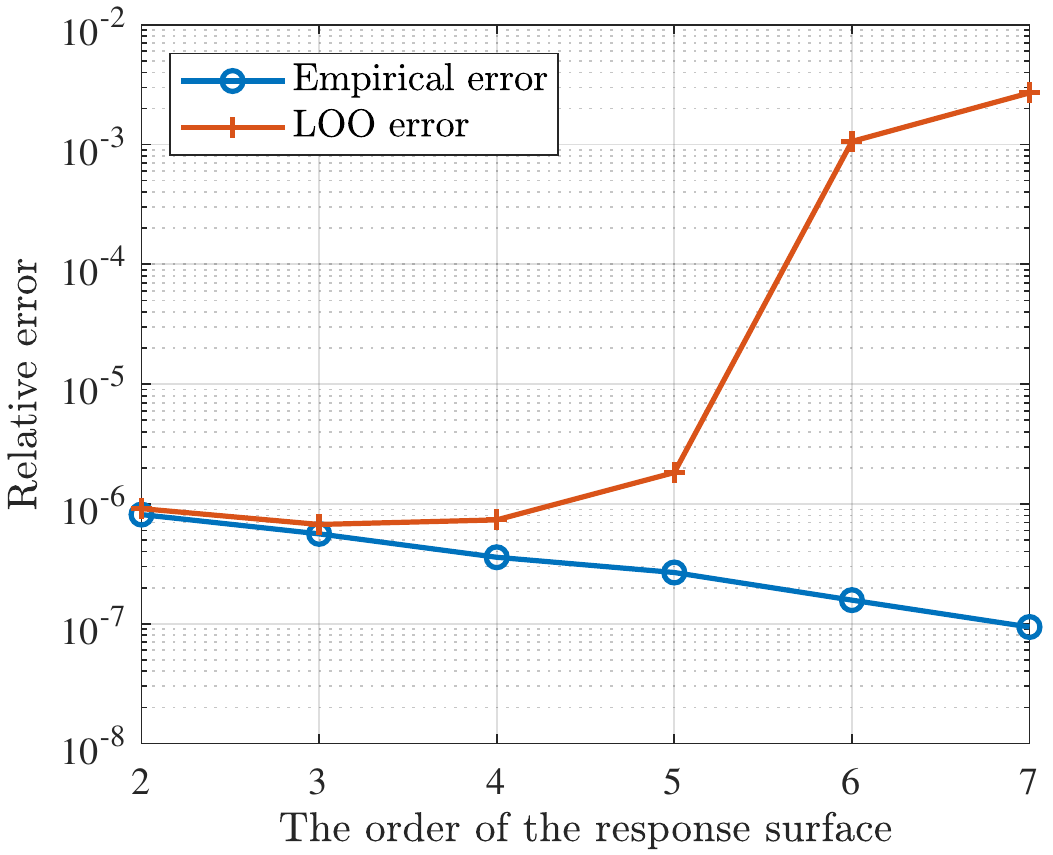}} 
	\subcaptionbox{The relative errors of $\wh{h}_2(\bm{\theta})$ \label{fig:lsr_err_pm}}
	{\includegraphics[width=.464\textwidth]{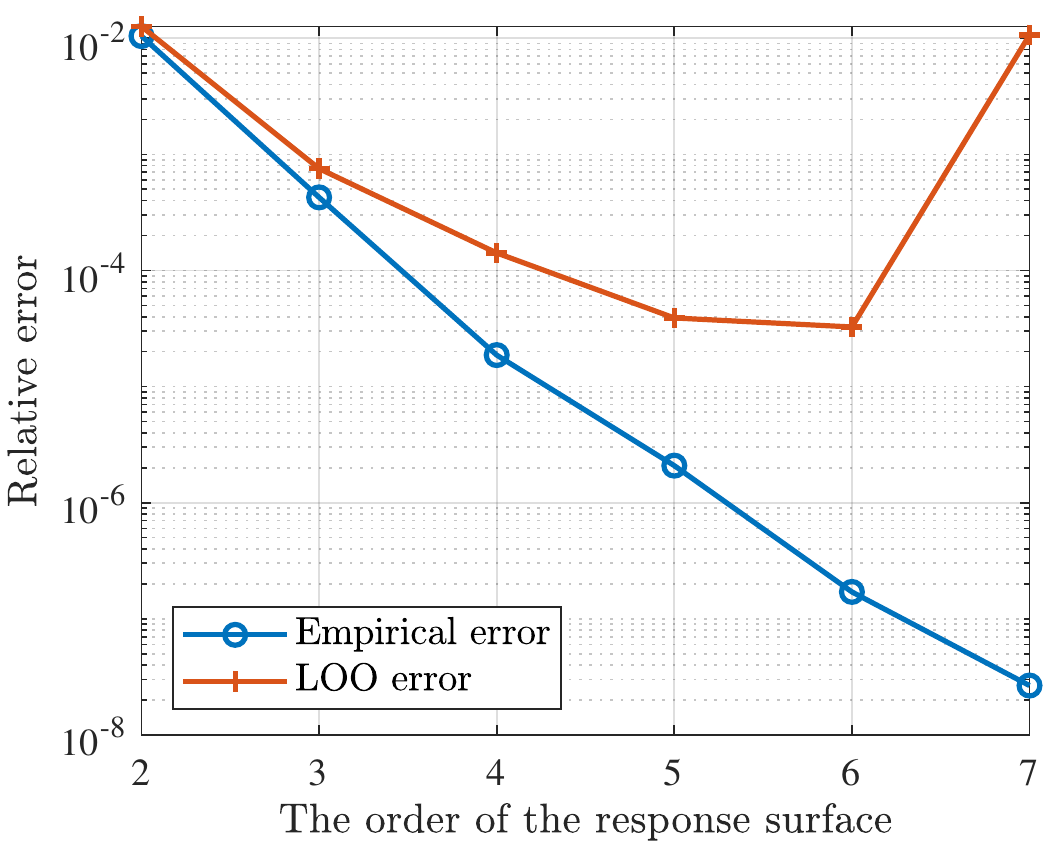}} 	
	\caption{The relative errors of the surrogate models.}
	\label{fig:lsr_err}
\end{figure}

Figure~\ref{fig:lsr_err} shows the qualities of the surrogate models constructed by RSM of different orders. The two subplots corresponding to $\wh{h}_1(\bm{\theta})$ and $\wh{h}_2(\bm{\theta})$ at a certain level of SBSS illustrate that the empirical error underestimates the fitting errors especially when excessively high orders are adopted. The surrogate model with the minimal LOO error is obtained by the adaptive RSM and the overfitting problem is thus avoided.

\begin{figure}[t!]
	\centering
	\subcaptionbox{The results of $\wh{h}_1(\bm{\theta})$ \label{fig:looerr_gm}}
	{\includegraphics[width=.460\textwidth]{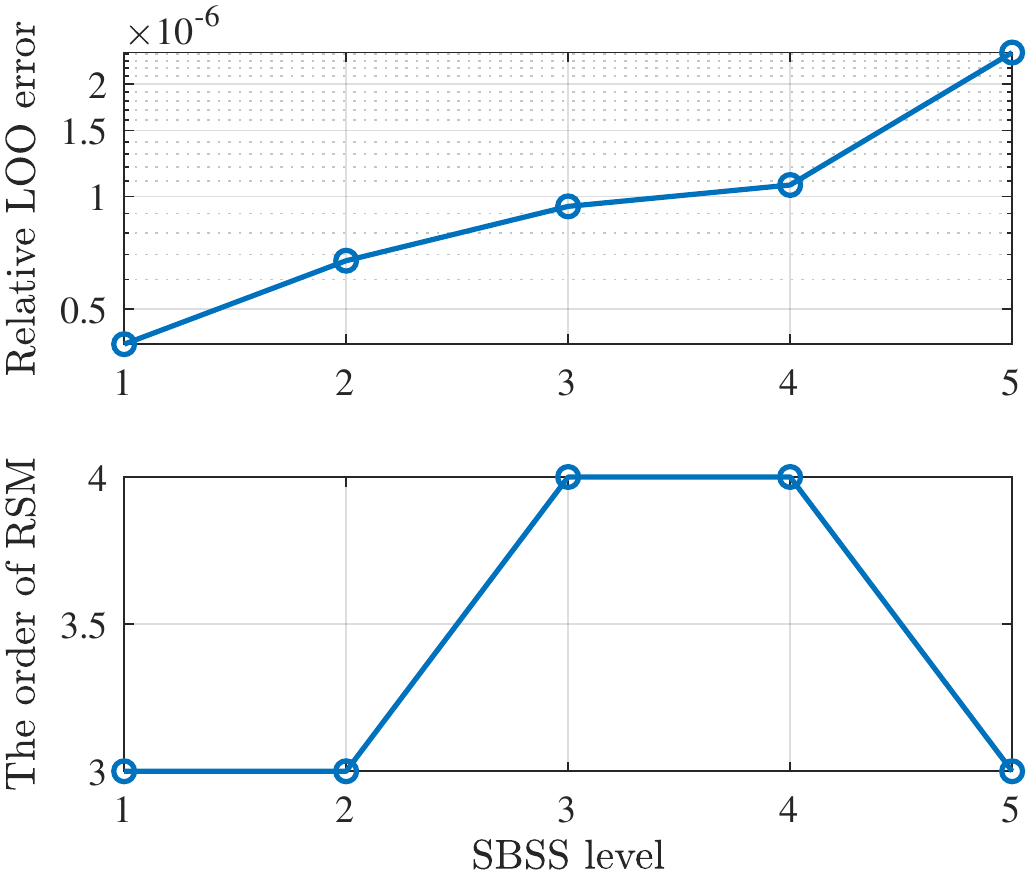}} 
	\subcaptionbox{The results of $\wh{h}_2(\bm{\theta})$ \label{fig:looerr_pm}}
	{\includegraphics[width=.468\textwidth]{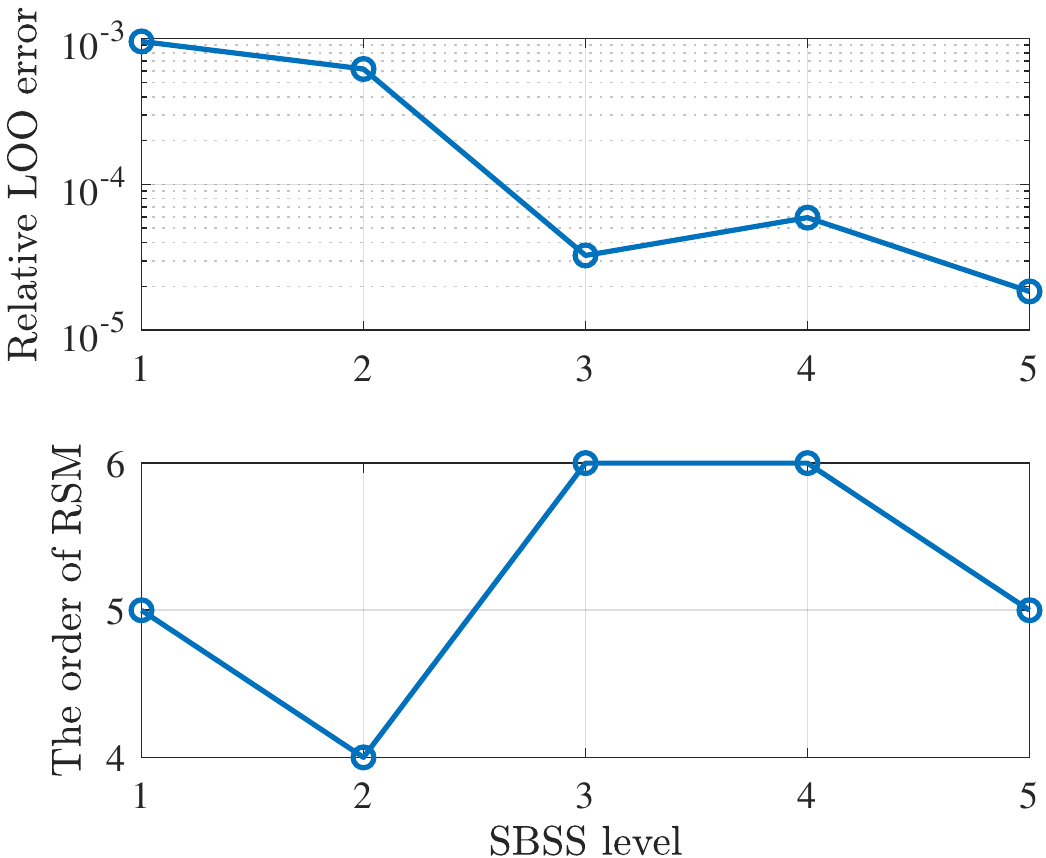}} 	
	\caption{The orders of RSM and the relative LOO errors.}
	\label{fig:looerr}
\end{figure}

\begin{figure}[t!]
	\centering
	\subcaptionbox{$h_1({\bm{\theta}})$ vs. $\wh{h}_1(\bm{\theta})$ \label{fig:tmsm_gm}}
	{\includegraphics[width=.415\textwidth]{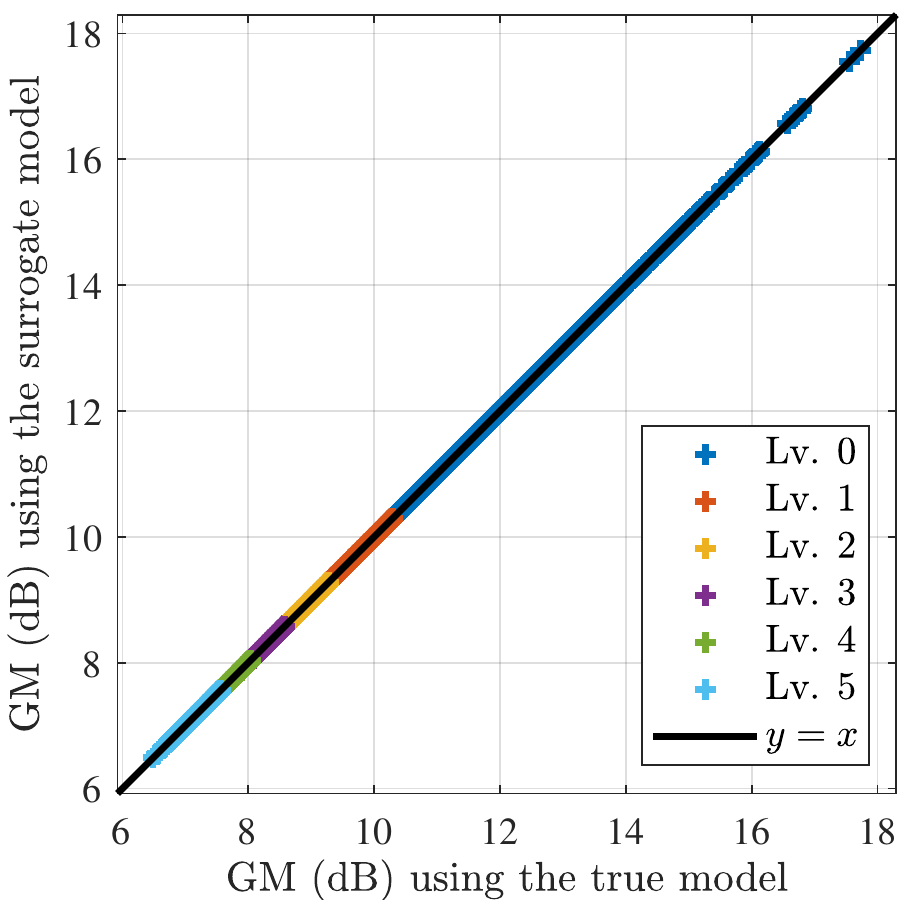}}
	\subcaptionbox{$h_2({\bm{\theta}})$ vs. $\wh{h}_2(\bm{\theta})$ \label{fig:tmsm_pm}}
	{\includegraphics[width=.422\textwidth]{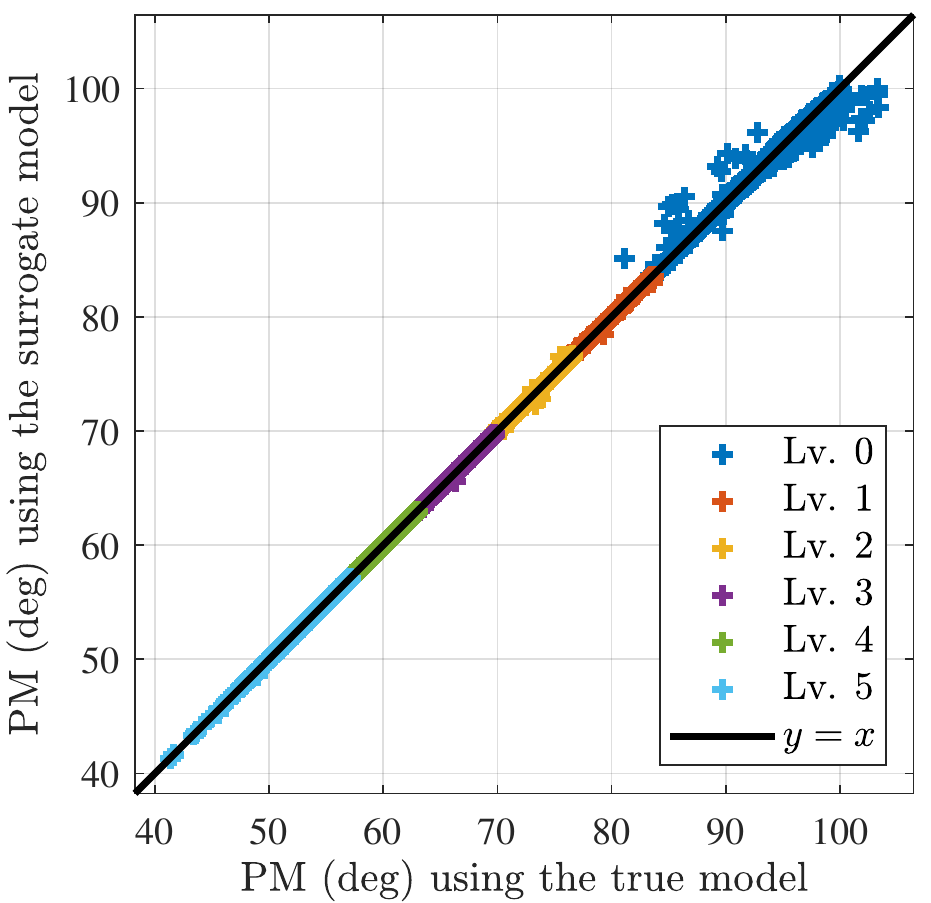}}
	\caption{Response comparisons (TM vs. SM by SBSS).}
	\label{fig:tmsm}
\end{figure}

\begin{figure}[t!]
	\centering
	\subcaptionbox{The results of gain margin \label{fig:cdf_gm}}
	{\includegraphics[width=.460\textwidth]{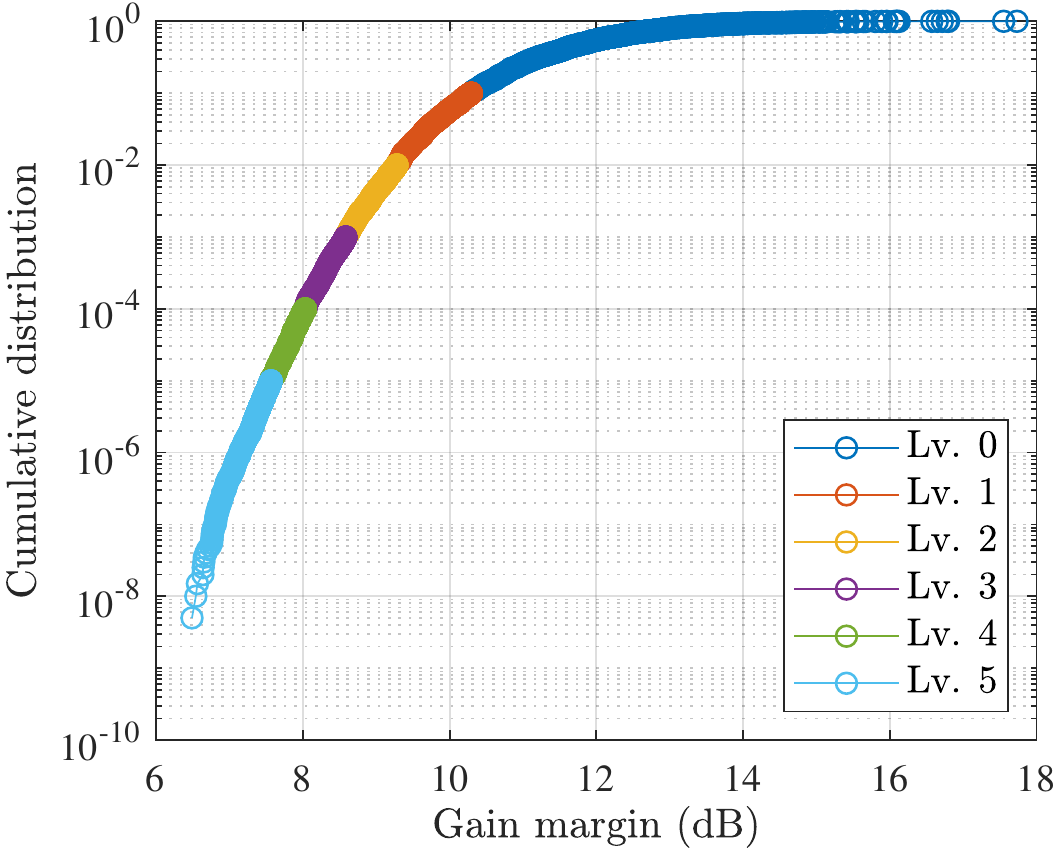}}
	\subcaptionbox{The results of phase margin \label{fig:cdf_pm}}
	{\includegraphics[width=.464\textwidth]{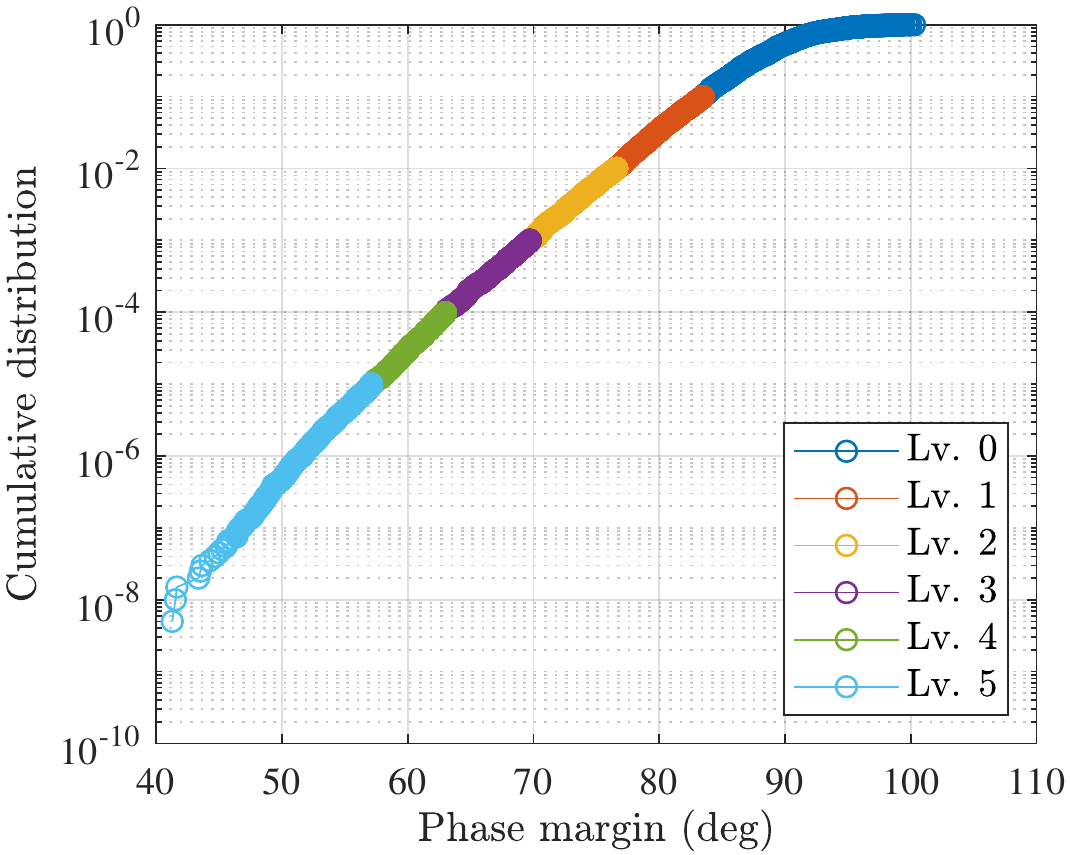}}
	\caption{The estimations of CDFs using SBSS.}
	\label{fig:cdf}
\end{figure}

The results of SBSS (level $0, 1, \ldots, 5$) are given in Figs.~\ref{fig:looerr}, \ref{fig:tmsm}, and \ref{fig:cdf}. Figure~\ref{fig:looerr} depicts the adopted orders of the adaptive RSM and the LOO errors of the built response surfaces. The subfigures show that the performance function $h_1(\bm{\theta})$ can be approximated by a low ($3$rd or $4$th) order polynomial whereas a relatively high ($5$th or $6$th) order polynomial may be required to describe the nonlinearity of $h_2(\bm{\theta})$.

\begin{figure}[t!]
	\centering
	\subcaptionbox{$h_1({\bm{\theta}})$ vs. $\wh{h}_1(\bm{\theta})$ \label{fig:tmsm_gm10}}
	{\includegraphics[width=.415\textwidth]{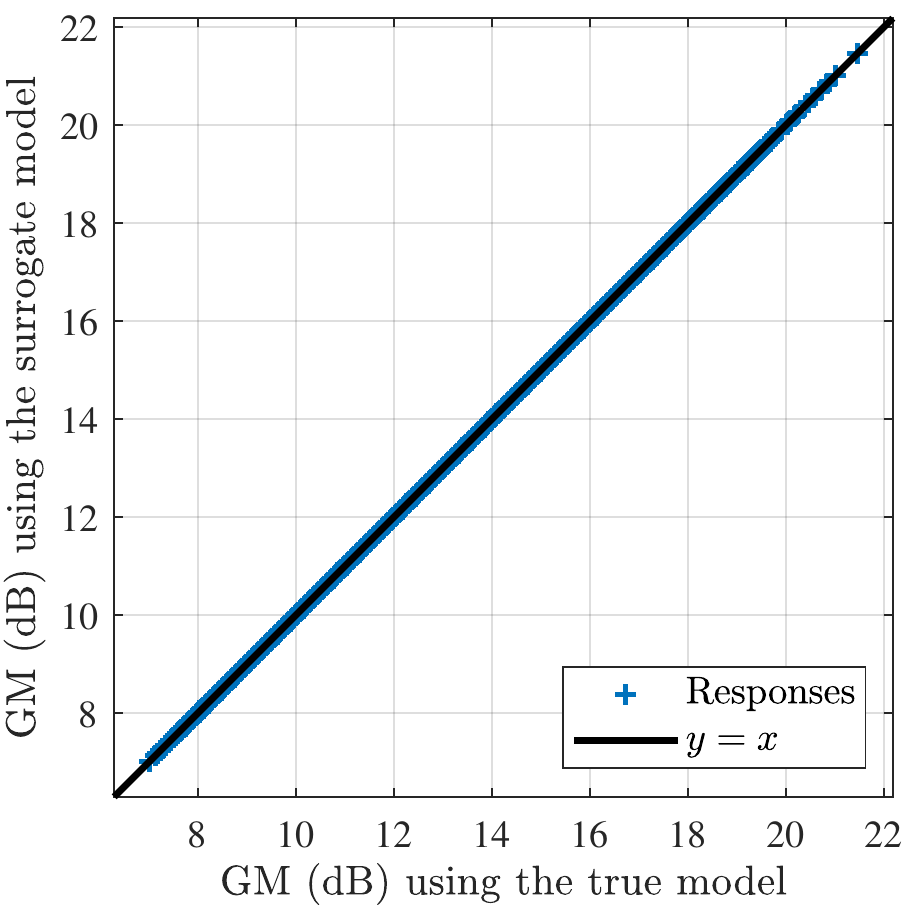}} 
	\subcaptionbox{$h_2({\bm{\theta}})$ vs. $\wh{h}_2(\bm{\theta})$ \label{fig:tmsm_pm10}}
	{\includegraphics[width=.422\textwidth]{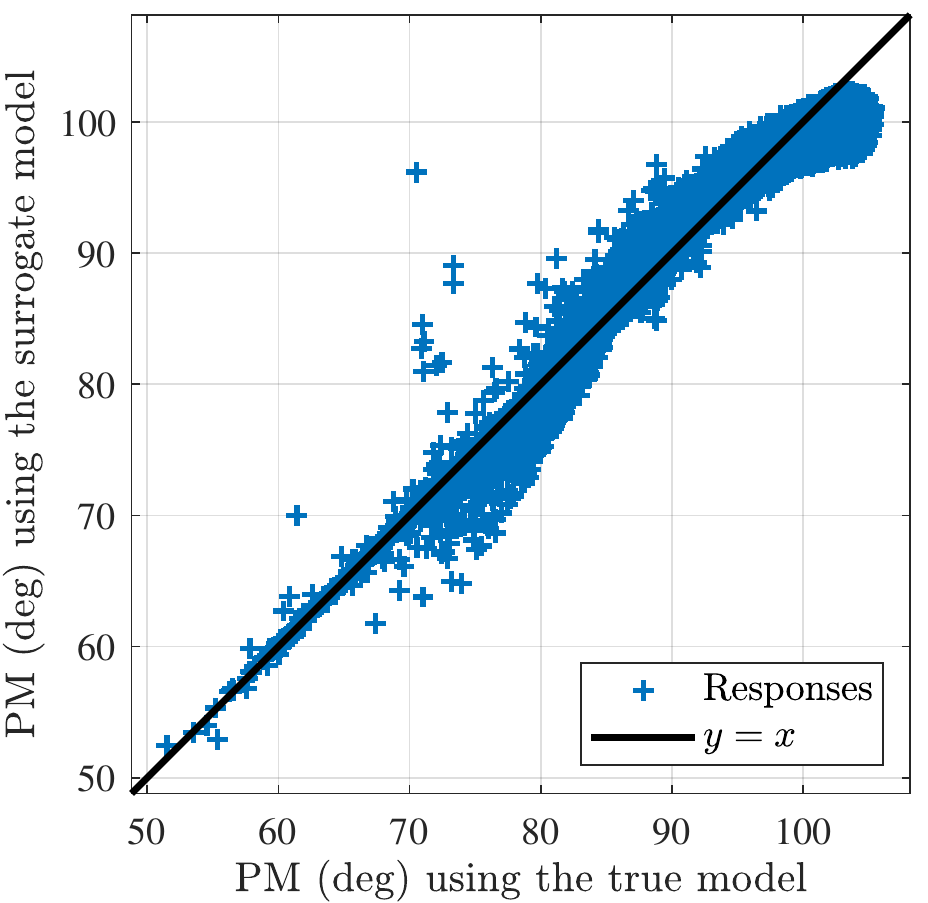}}	
	\caption{Response comparisons (TM vs. SM by PCE).}
	\label{fig:tmsm10}
\end{figure}

In Fig.~\ref{fig:tmsm}, the responses of the true model (TM) and the surrogate model (SM) are compared given the same uncertain inputs. This comparison is also conducted between the responses of the true model and the surrogate model constructed by the $10$th-order PCE (spectral projection) in Fig.~\ref{fig:tmsm10}. In these two figures, the closer these points lie to the line $y=x$, the better the surrogates. The noncompliance with the reference line in Fig.~\ref{fig:tmsm_pm10} indicates that the performance function $h_2({\bm{\theta}})$ can hardly be represented by a polynomial even though the order is as high as $10$. However, it can be solved by a relatively low order piecewise polynomial surrogate as shown in Fig.~\ref{fig:tmsm_pm}. This implies that $h_2({\bm{\theta}})$ might be highly nonlinear or even not smooth, and the nonlinearity mainly influence the surrogate model at the initial level. Despite the defective initial surrogate model, further ones towards the failure domain are constructed successfully. Therefore, building a piecewise surrogate model could be an effective way to approximate a highly nonlinear function. On the contrary, $h_1({\bm{\theta}})$ can be expressed as a polynomial function since all the samples in Fig.~\ref{fig:tmsm_gm10} are perfectly close to the line $y=x$.

\begin{figure}[t!]
	\centering
	\subcaptionbox{The results of gain margin \label{fig:cdf_pf_gm}}
	{\includegraphics[width=.460\textwidth]{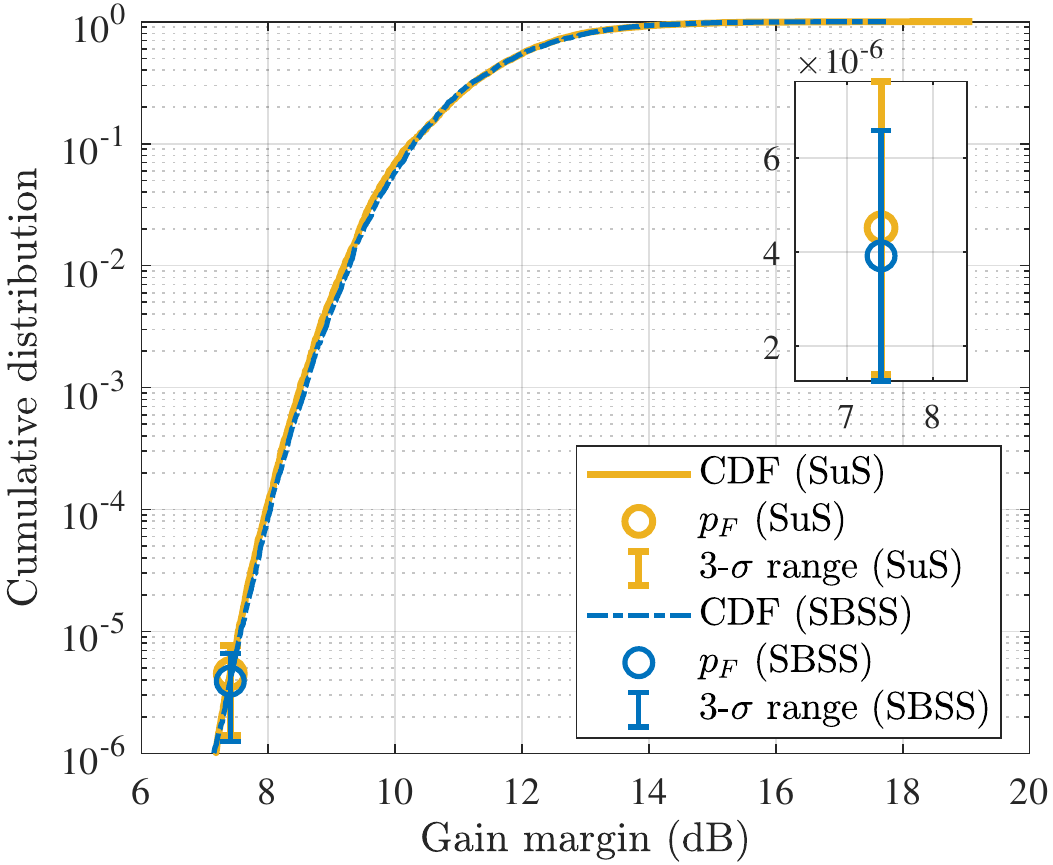}} 
	\subcaptionbox{The results of phase margin \label{fig:cdf_pf_pm}}
	{\includegraphics[width=.463\textwidth]{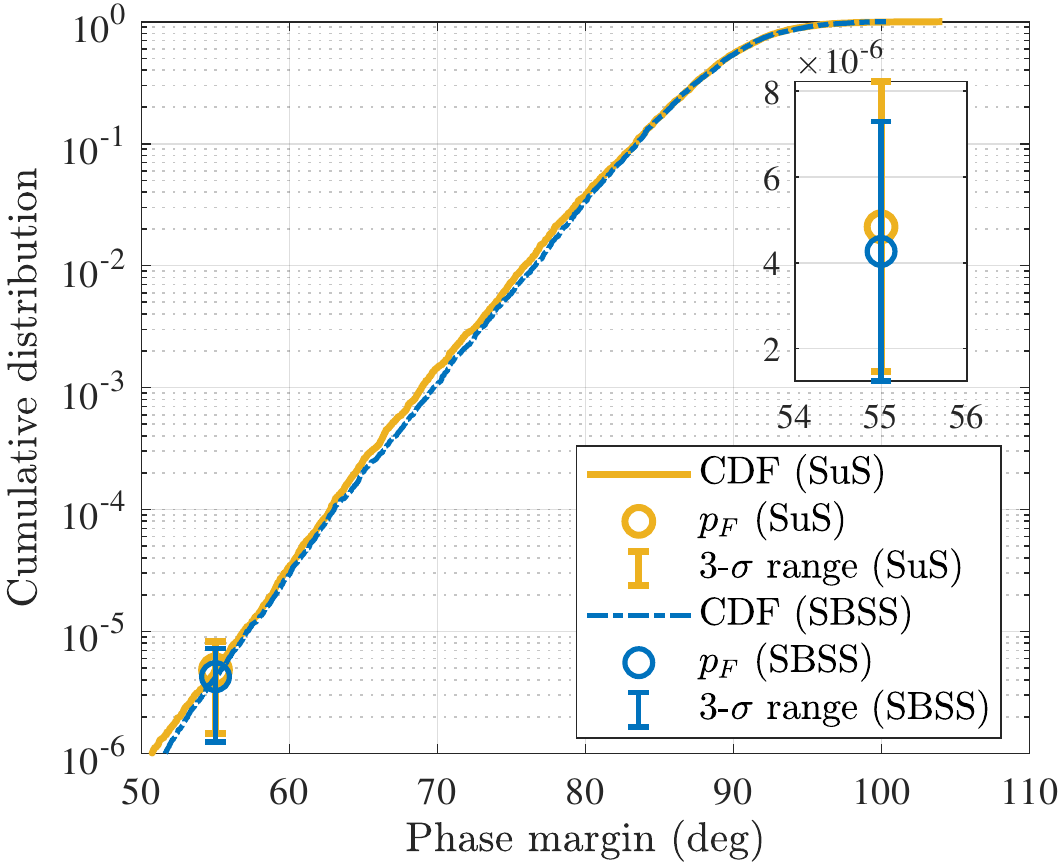}} 
	\caption{The estimations of CDFs and failure probabilities.}
	\label{fig:cdf_pf}
\end{figure}

\begin{figure}[t!]
	\centering
	\subcaptionbox{The results of gain margin \label{fig:cov1_gm}}
	{\includegraphics[width=.420\textwidth]{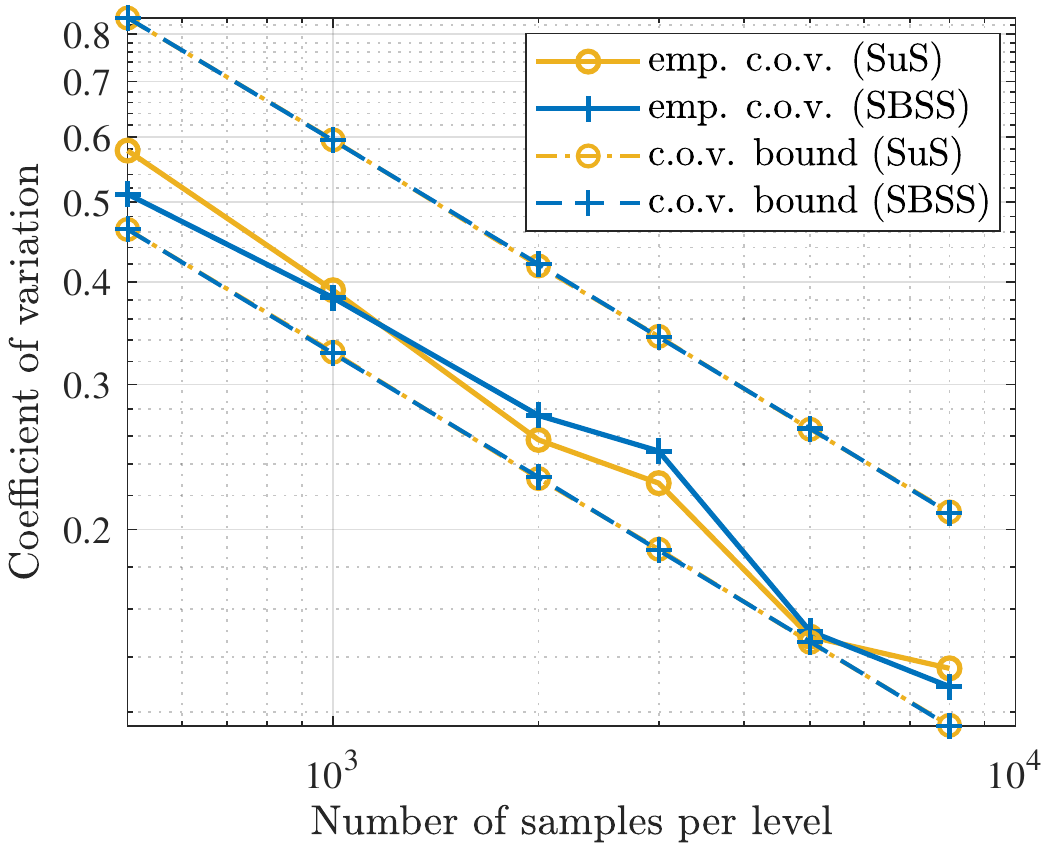}} 
	\subcaptionbox{The results of phase margin \label{fig:cov1_pm}}
	{\includegraphics[width=.420\textwidth]{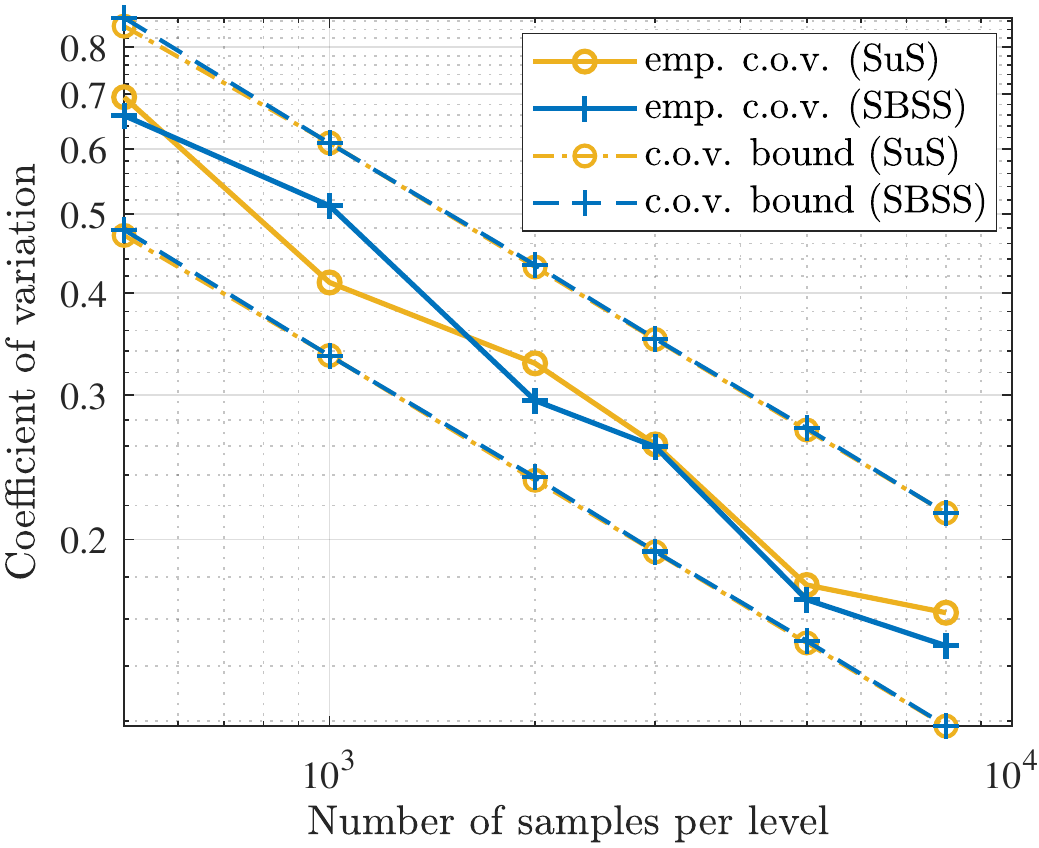}} 	
	\caption{The estimations of coefficients of variation.}
	\label{fig:cov1}
\end{figure}

\begin{figure}[t!]
	\centering
	\subcaptionbox{The results of gain margin \label{fig:pf1_gm}}
	{\includegraphics[width=.420\textwidth]{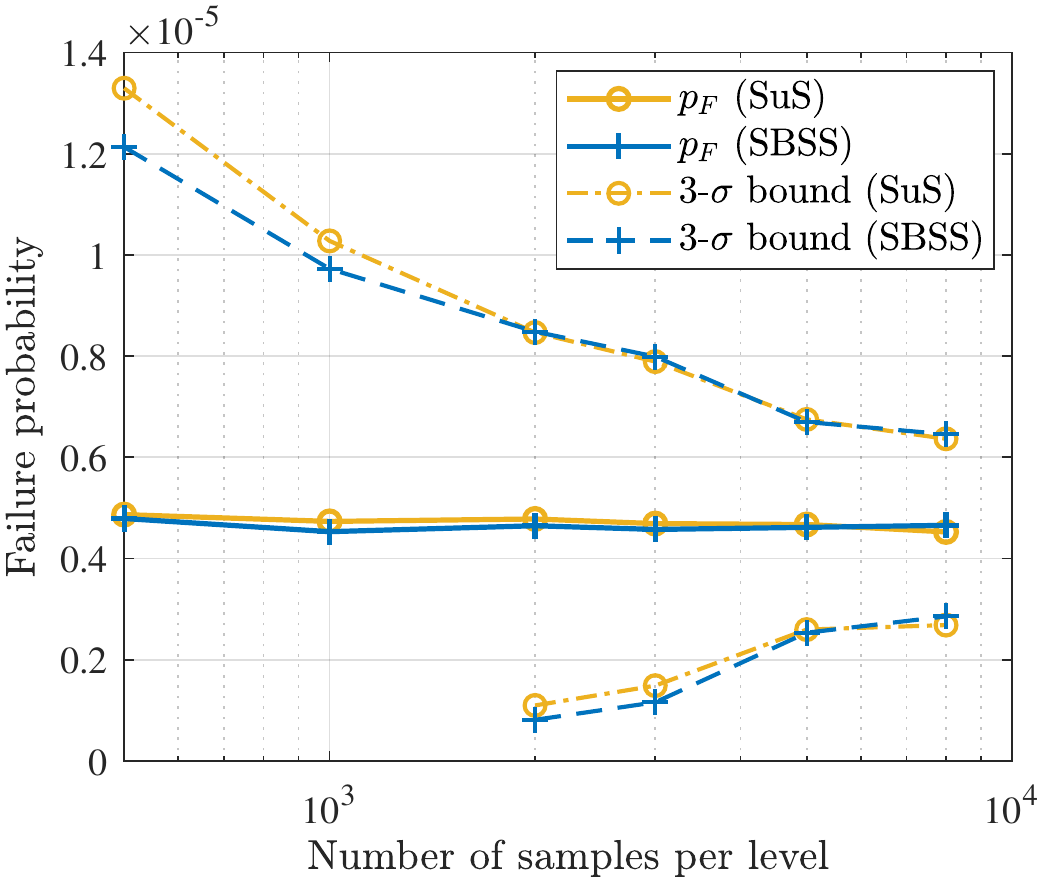}} 
	\subcaptionbox{The results of phase margin \label{fig:pf1_pm}}
	{\includegraphics[width=.420\textwidth]{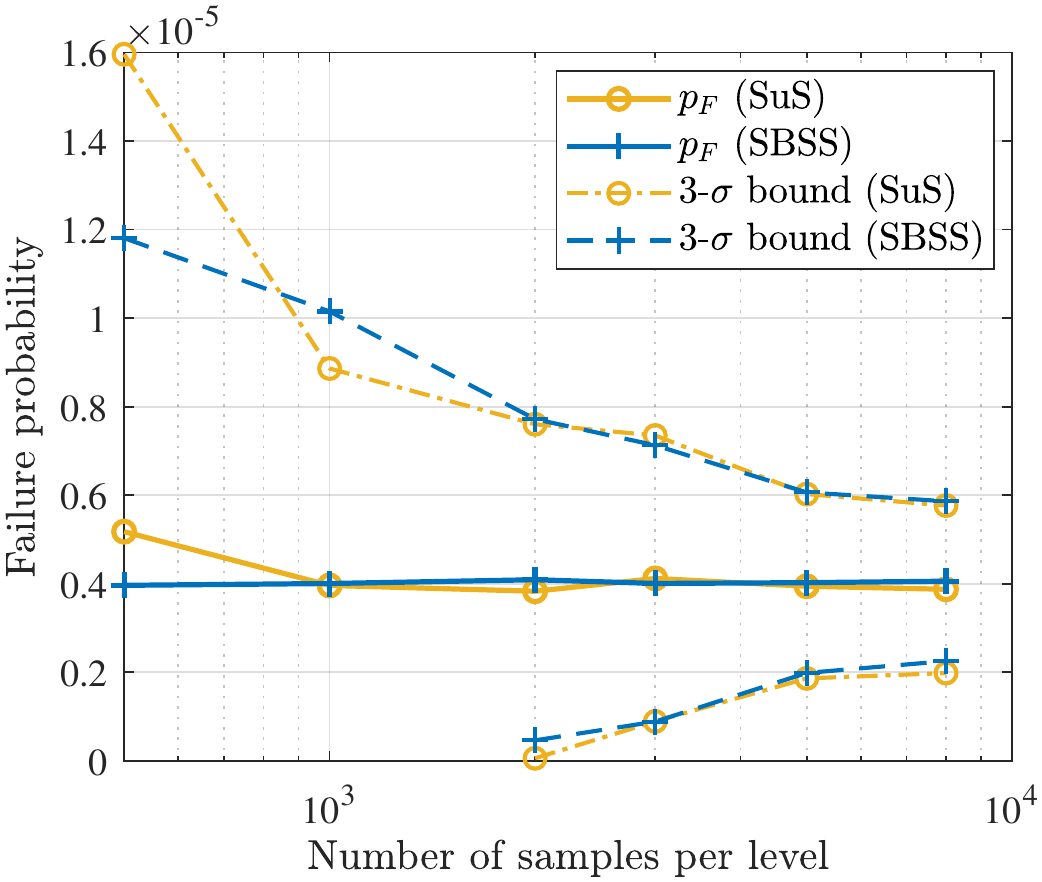}} 	
	\caption{The estimations of failure probabilities.}
	\label{fig:pf1}
\end{figure}

\begin{figure}[t!]
	\centering
	\subcaptionbox{The results of gain margin \label{fig:cov2_gm}}
	{\includegraphics[width=.419\textwidth]{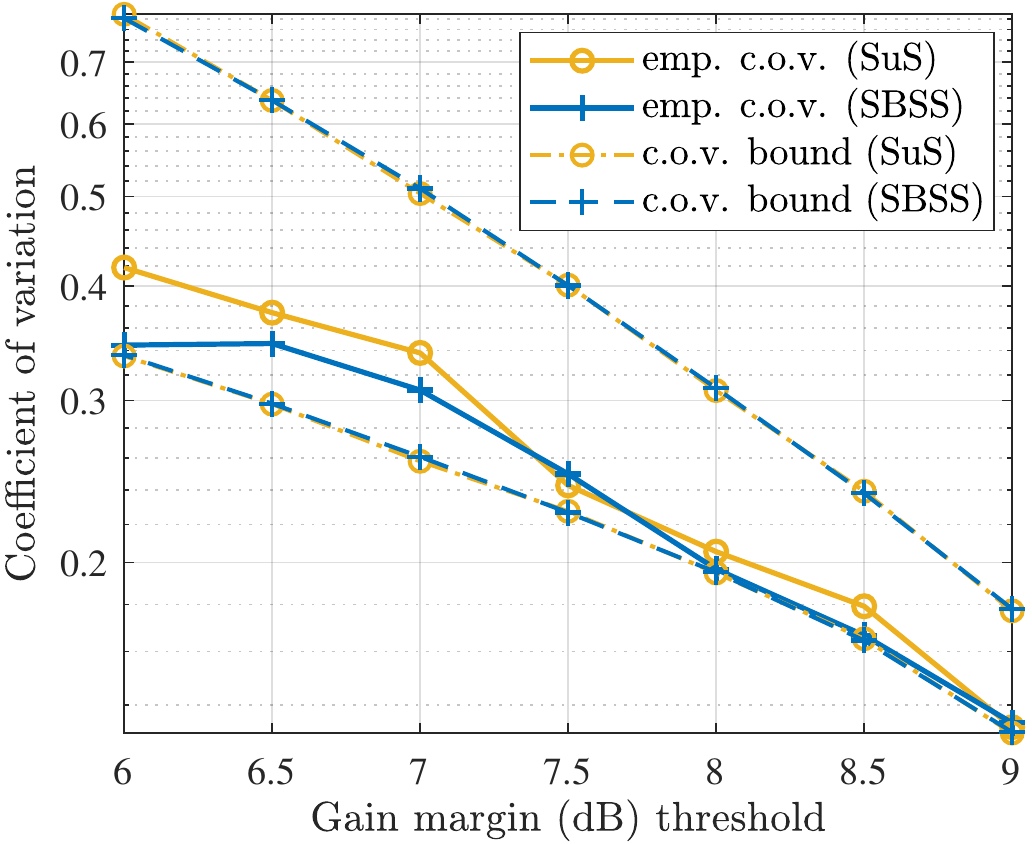}} 
	\subcaptionbox{The results of phase margin \label{fig:cov2_pm}}
	{\includegraphics[width=.425\textwidth]{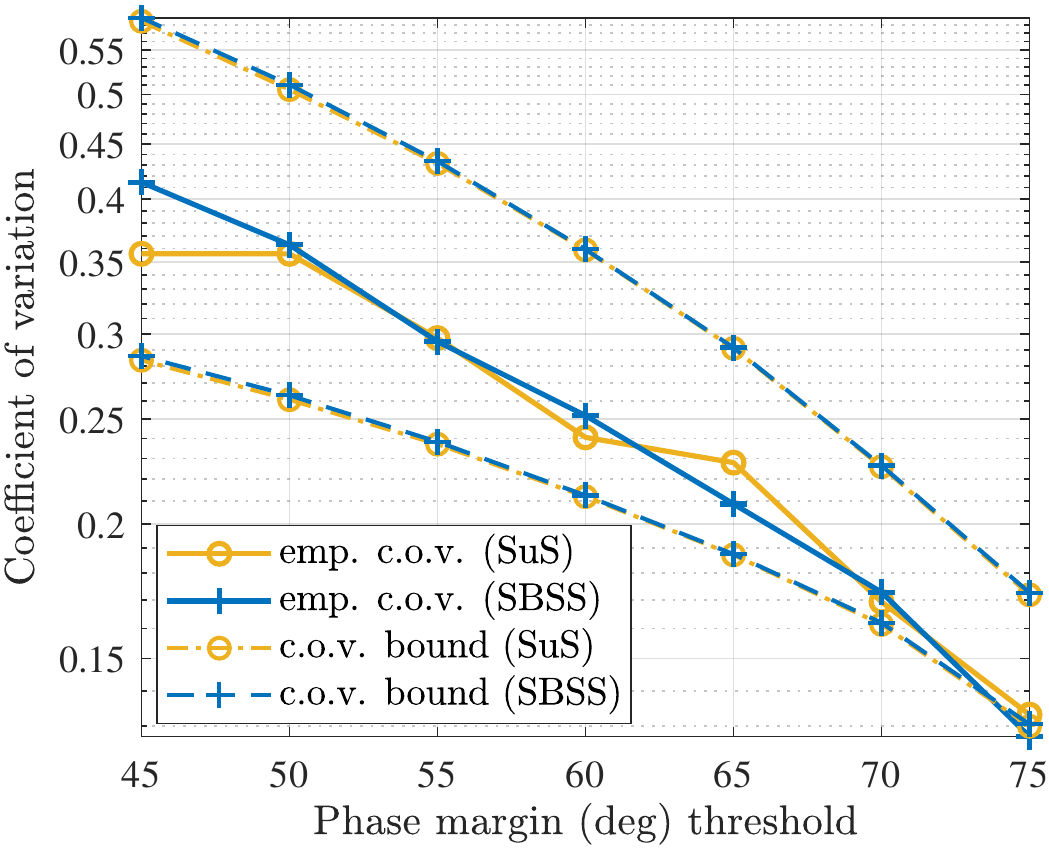}} 	
	\caption{The estimations of coefficients of variation.}
	\label{fig:cov2}
\end{figure}

\begin{figure}[t!]
	\centering
	\subcaptionbox{The results of gain margin \label{fig:pf2_gm}}
	{\includegraphics[width=.440\textwidth]{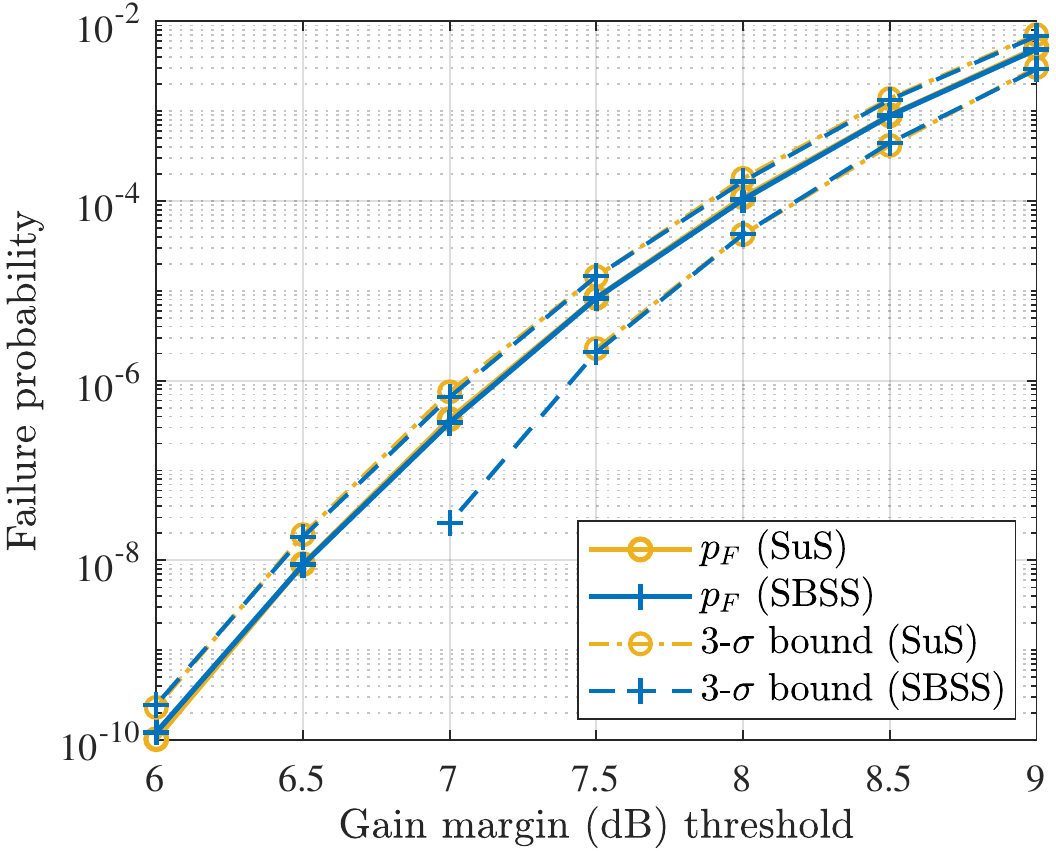}} 
	\subcaptionbox{The results of phase margin \label{fig:pf2_pm}}
	{\includegraphics[width=.440\textwidth]{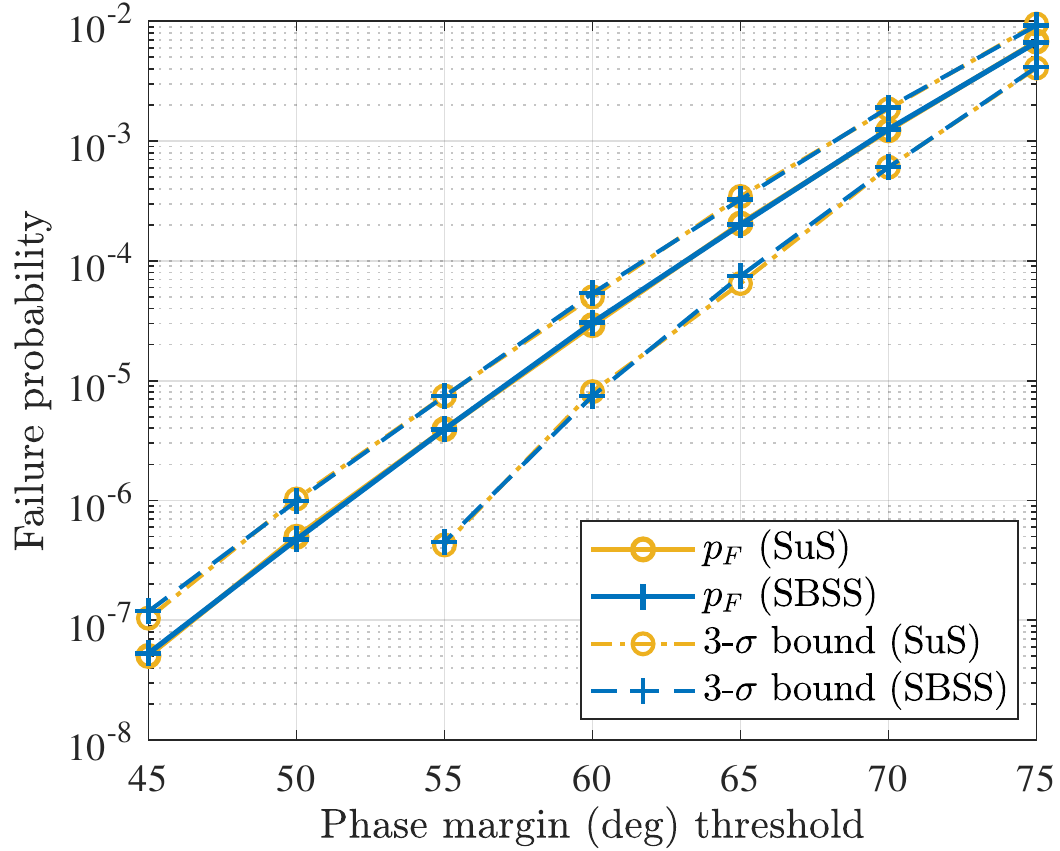}} 	
	\caption{The estimations of failure probabilities.}
	\label{fig:pf2}
\end{figure}

The results of cumulative distribution function (CDF) are depicted in Fig.~\ref{fig:cdf}. The samples in SBSS approach the increasingly rare domain level by level, which behave the same as those in SuS. Besides the CDFs, the failure probabilities of events $h_{1}(\bm{\theta}) < 7.4~\rm dB$ and $h_{2}(\bm{\theta}) < 55 ^{\circ}$ are also assessed in Fig.~\ref{fig:cdf_pf}. With 80 replicate assessments, Figures~\ref{fig:cov1} and~\ref{fig:pf1} illustrate the estimations of c.o.v. and $p_f$ using different numbers of samples at each simulation level. Here, the c.o.v. bound is the averaged estimates of c.o.v. upper or lower bounds \cite{Au2001} over 80 simulations, whereas the empirical (emp.) c.o.v. is the sample c.o.v. of the failure probability estimates over these independent runs. The $3$-$\sigma$ range is considered to cover nearly all the possible failure probabilities. It is estimated by the lower bound of c.o.v. in Fig.~\ref{fig:cdf_pf} and by the emp. c.o.v. in Fig.~\ref{fig:pf1}. In these graphs, both the CDFs and the failure probabilities estimated by SBSS are in great accordance with their counterparts by SuS, demonstrating the comparable performance of SBSS. Similar comparisons are conducted for different thresholds (2000 samples at each level of SuS and SBSS) in Figs.~\ref{fig:cov2} and \ref{fig:pf2}, and the matches of the results also prove that SBSS owns the equivalent capability of estimating rare-event probabilities.

The numbers of calls to the true model (denoted by $N_{\mathrm{call}}$) using SuS and SBSS are depicted in Tables~\ref{table:Ntm1} and \ref{table:Ntm2}. Table~\ref{table:Ntm1} shows the numbers of true model evaluations with different numbers of samples per level $N$ and Table~\ref{table:Ntm2} shows those with different numbers of total levels $m$. In all cases, at least $86\%$ of the calls to the true model are saved by SBSS, demonstrating its superior efficiency over SuS.

\begin{table}[t!]
	\centering
	\caption{The numbers of calls to the true model with different number of samples per level.}\label{table:Ntm1}
	\begin{tabular}{cccc}
		\hline
		$N$ & $m$ & $N_{\mathrm{call}}$ (SuS) & $N_{\mathrm{call}}$ (SBSS) \\ \hline
		1000 & 6 &  5500 & 766  \\
		2000 & 6 & 11000 & 1316 \\
		3000 & 6 & 16500 & 1866 \\
		5000 & 6 & 27500 & 2966 \\ \hline
	\end{tabular}
\end{table}

\begin{table}[t!]
	\centering
	\caption{The numbers of calls to the true model with different numbers of total levels.}\label{table:Ntm2}
	\begin{tabular}{cccc}
		\hline
		$N$ & $m$ & $N_{\mathrm{call}}$ (SuS) & $N_{\mathrm{call}}$ (SBSS) \\ \hline
		2000 & 2 &  3800 & 436  \\
		2000 & 4 &  7400 & 876  \\
		2000 & 6 & 11000 & 1316 \\
		2000 & 8 & 14600 & 1756 \\
		2000 & 10 & 18200 & 2196 \\ \hline
	\end{tabular}
\end{table}

\subsection{Control Parameter Optimization}

Besides the stability margin requirements in Eq.~\eqref{eq:sm}, tracking performance and disturbance rejection performance are considered as well in the optimization framework Eq.~\eqref{eq:opt}. The objective function and the probabilistic constraints are defined in Table~\ref{table:opt}.

\begin{table}[b!]
	\centering
	\caption{The objective function and the probabilistic constraints.}\label{table:opt}
	\begin{tabular}{ccc}
		\hline
		\makecell{Performance\\functions} & Descriptions & \makecell{Corresponding objective function\\or constraints} \\ \hline
		$h_0(\bm{\theta}, \bm{k})$ & Maximum negative deviation of gust reaction & $\mathop{\mathbb{P}}[h_0(\bm{\theta}, \bm{k}) < -0.45]$  \\
		$h_1(\bm{\theta}, \bm{k})$ & Gain margin & $\mathop{\mathbb{P}}[h_1(\bm{\theta}, \bm{k}) < 6~\rm dB] < 10^{-6}$ \\
		$h_2(\bm{\theta}, \bm{k})$ & Phase margin & $\mathop{\mathbb{P}}[h_2(\bm{\theta}, \bm{k}) < 45 ^{\circ}] < 10^{-6}$ \\
		$h_3(\bm{\theta}, \bm{k})$ & Overshoot of step response & $\mathop{\mathbb{P}}[h_3(\bm{\theta}, \bm{k}) > 20\%] < 0.1$ \\
		$h_4(\bm{\theta}, \bm{k})$ & $80\%$ rise time of step response & $\mathop{\mathbb{P}}[h_4(\bm{\theta}, \bm{k}) > 1~\rm s] < 0.1$ \\ \hline
	\end{tabular}
\end{table}

Given the mean values of the uncertain parameters, the deterministic constraint functions are defined as
\begin{equation} \label{eq:opt1}
\begin{array}{l}
c_1(\bm{k}) = \dfrac{1}{T} \displaystyle{ \int_{0}^{T} \ddot{q}_{\rm cmd}^2 } \, \mathrm{d} t - C_{d,1}, \\
c_2(\bm{k}) = \dfrac{1}{T} \displaystyle{ \int_{0}^{T} \dot{n}_z^2 } \, \mathrm{d} t - C_{d,2}, \\
\end{array}
\end{equation}
where $T$ is the simulation time, $C_{d,1}$ and $C_{d,2}$ are the thresholds for the mean square values. $c_1(\bm{k})$ and $c_2(\bm{k})$ limit the oscillation of control input $\dot{q}_{\rm cmd}$ and system output $n_z$.

The rare-event probabilities are estimated by the proposed SBSS method ($2000$ samples at each level) while the others are approximated by MCS ($10^5$ samples) based on the surrogate model constructed by the 5th-order PCE. The expansion coefficients are computed employing the spectral projection with a tensor product quadrature (6 nodes in each dimension and 216 nodes in total). To guarantee the fulfillment of the rare-event requirements, the $3$-$\sigma$ upper bounds are considered to satisfy the constraints:
\begin{equation} \label{eq:opt2}
\begin{array}{ll}
(1+3c_{v,1})\mathop{\mathbb{P}}[h_{1}(\bm{\theta}, \bm{k}) < 6~\rm dB] < 10^{-6}, \\
(1+3c_{v,2})\mathop{\mathbb{P}}[h_{2}(\bm{\theta}, \bm{k}) < 45 ^{\circ}] < 10^{-6},
\end{array}
\end{equation}
where $c_{v,1}$ and $c_{v,2}$ are the c.o.v. of the estimates of $\mathop{\mathbb{P}}[h_{1}(\bm{\theta}) < 6~\rm dB]$ and $\mathop{\mathbb{P}}[h_{2}(\bm{\theta}) < 45 ^{\circ}]$ respectively. The c.o.v. can be reduced by increasing the number of samples at each level of SBSS. This, however, increases the computational burden. Therefore, the choice of the sample number is a tradeoff between the conservativeness of constraint bounds and the computational cost.

\begin{figure}[t!]
	\centering
	\subcaptionbox{The probabilistic bounds of gust reaction \label{fig:valid_gust}}
	{\includegraphics[width=.440\textwidth]{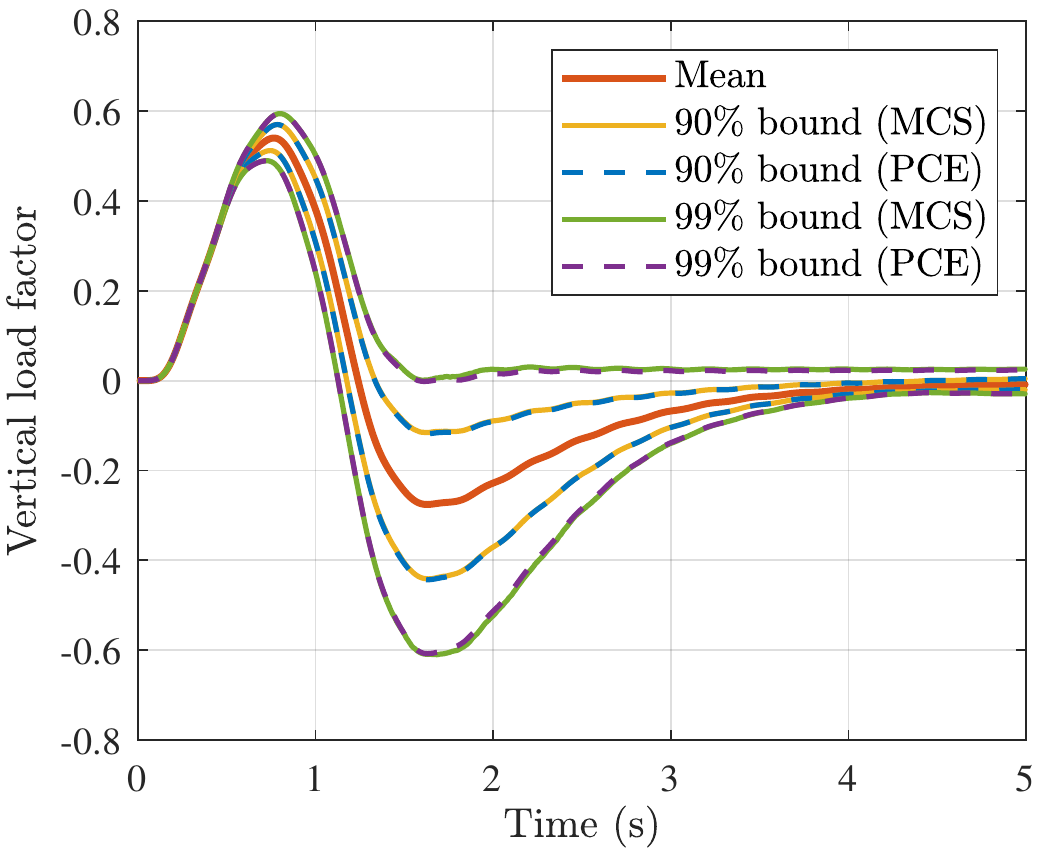}} 
	\subcaptionbox{The probabilistic bounds of step response \label{fig:valid_step}}
	{\includegraphics[width=.440\textwidth]{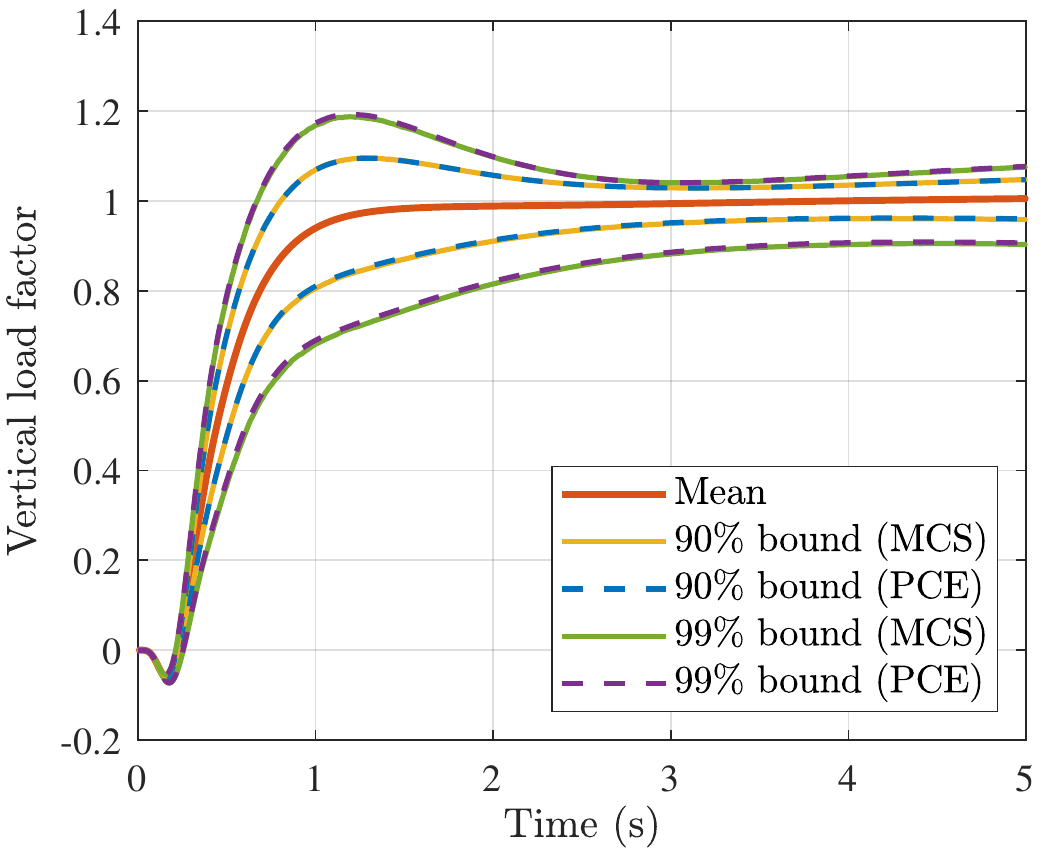}} 	
	\caption{The estimations of probabilistic bounds.}
	\label{fig:valid}
\end{figure}

The optimization problem is solved by the Matlab global solver \emph{surrogateopt} \cite{surrogateopt}. Given the designed control gains, the violation probabilities of performance functions $p_v$ and the numbers of calls to the true model $N_{\mathrm{call}}$ are shown in Table~\ref{table:opt_result}. The results of MCS ($10^4$ samples) and SuS are regarded as references. The probabilistic bounds of gust reaction and step response are depicted in Fig.~\ref{fig:valid}. As given in Table~\ref{table:opt_result} and Fig.~\ref{fig:valid}, the chance constraints in Table~\ref{table:opt} are fulfilled directly during the minimization of the object function. Meanwhile, the results of PCE and SBSS match well with those of MCS and SuS, but far fewer true model evaluations are required in these surrogate-based methods. This suggests that the surrogate-based methods are able to provide reasonably accurate estimates with much higher efficiency for both failure probability estimation and chance-constrained optimization.

\begin{table}[t!]
	\centering
	\caption{The violation probabilities of performance functions.}\label{table:opt_result}
	\begin{tabular}{ccc}
		\hline
		\makecell{Violation\\probabilities} & \makecell{MCS or SuS\\$p_v$ ($N_{\mathrm{call}}$)} & \makecell{PCE or SBSS\\$p_v$ ($N_{\mathrm{call}}$)} \\ \hline
		$\mathop{\mathbb{P}}[h_0(\bm{\theta}, \bm{k}) < -0.45]$ & $0.0953$ ($10000$)& $0.0947$ ($216$)  \\
		$\mathop{\mathbb{P}}[h_1(\bm{\theta}, \bm{k}) < 6~\rm dB]$ & $2.71\times10^{-7}$, $c_v\in[0.27, 0.52]$ ($12800$) & $2.04\times10^{-7}$, $c_v\in[0.26, 0.50]$ ($1536$) \\
		$\mathop{\mathbb{P}}[h_2(\bm{\theta}, \bm{k}) < 45 ^{\circ}]$ & $<1\times10^{-10}$ ($18200$) & $<1\times10^{-10}$ ($2196$) \\
		$\mathop{\mathbb{P}}[h_3(\bm{\theta}, \bm{k}) > 20\%]$ & $0.0082$ ($10000$) & $0.0087$ ($216$) \\
		$\mathop{\mathbb{P}}[h_4(\bm{\theta}, \bm{k}) > 1~\rm s]$ & $0.0872$ ($10000$) & $0.0865$ ($216$) \\ \hline
	\end{tabular}
\end{table}

\section{Conclusions}

A surrogate-based subset simulation (SBSS) method is developed and applied to estimate rare-event probabilities in the framework of performance-guaranteed control optimization. Compared with conventional subset simulation (SuS), the proposed SBSS method saves a large number of calls to the true model, thus gaining higher efficiency for both rare failure probability approximation and rare-event chance-constrained optimization. The simulation results demonstrate that SBSS provides a comparable performance to SuS but with much less computational expense and the performance-based control optimization directly ensures the fulfillment of requirements. We plan to extend this concept to solve high-dimensional problems in the future work.


\section*{Acknowledgments}
The first author acknowledges the financial support from China Scholarship Council (CSC) on his doctoral study at TUM.

\bibliography{sample}

\end{document}